\documentclass[aps,pra,reprint,superscriptaddress,floatfix]{revtex4-2}

\usepackage{amsmath}
\usepackage{amstext}
\usepackage{amsfonts}
\usepackage{multirow}
\usepackage{pifont}
\usepackage{xspace}
\usepackage{graphicx}
\usepackage{dcolumn}
\usepackage{bm}
\usepackage{color}
\usepackage{units}

\newcommand{\cmark}{\ding{51}}%
\renewcommand{\vec}[1]{\boldsymbol{#1}}

\newcommand{\equref}[1]{Eq.~(\ref{#1})}

\newcommand{\secref}[1]{Sec.~\ref{#1}}
\newcommand{\figref}[1]{Fig.~\ref{#1}}
\newcommand{\refcite}[1]{Ref.~\onlinecite{#1}}
\newcommand{\refscite}[1]{Refs.~\onlinecite{#1}}

\newcommand{\tableref}[1]{Table~\ref{#1}}
\newcommand{\appref}[1]{Appendix~\ref{#1}}
\newcommand{\pdagger}{{\phantom{\dagger}}}
\usepackage{braket}

\begin{document}
\title{Possible unconventional pairing in $(\text{Ca,Sr})_{3}(\text{Ir,Rh})_{4}\text{Sn}_{13}$\\
superconductors revealed by controlling disorder}

\author{E.~H.~Krenkel}
\affiliation{Ames Laboratory, Ames, Iowa 50011, USA}
\affiliation{Department of Physics and Astronomy, Iowa State University, Ames,
Iowa 50011, USA}

\author{M.~A.~Tanatar}
\affiliation{Ames Laboratory, Ames, Iowa 50011, USA}
\affiliation{Department of Physics and Astronomy, Iowa State University, Ames,
Iowa 50011, USA}

\author{M.~Ko\'{n}czykowski}
\affiliation{Laboratoire des Solides Irradi\'{e}s, CEA/DRF/lRAMIS, \'{E}cole Polytechnique, CNRS, Institut Polytechnique de Paris, F-91128 Palaiseau, France}

\author{R.~Grasset}
\affiliation{Laboratoire des Solides Irradi\'{e}s, CEA/DRF/lRAMIS, \'{E}cole Polytechnique, CNRS, Institut Polytechnique de Paris, F-91128 Palaiseau, France}

\author{E.~I.~Timmons}
\affiliation{Ames Laboratory, Ames, Iowa 50011, USA}
\affiliation{Department of Physics and Astronomy, Iowa State University, Ames,
Iowa 50011, USA}

\author{S.~Ghimire}
\affiliation{Ames Laboratory, Ames, Iowa 50011, USA}
\affiliation{Department of Physics and Astronomy, Iowa State University, Ames,
Iowa 50011, USA}

\author{K.~R.~Joshi}
\affiliation{Ames Laboratory, Ames, Iowa 50011, USA}
\affiliation{Department of Physics and Astronomy, Iowa State University, Ames,
Iowa 50011, USA}

\author{Y.~Lee}
\affiliation{Ames Laboratory, Ames, Iowa 50011, USA}

\author{Liqin~Ke}
\affiliation{Ames Laboratory, Ames, Iowa 50011, USA}

\author{S.~Chen}
\affiliation{Condensed Matter Physics and Materials Science Department, Brookhaven
National Laboratory, Upton, New York 11973, USA}
\affiliation{Department of Physics and Astronomy, Stony Brook University, Stony
Brook, NY 11794-3800, USA}

\author{C.~Petrovic}
\affiliation{Condensed Matter Physics and Materials Science Department, Brookhaven
National Laboratory, Upton, New York 11973, USA}
\affiliation{Department of Physics and Astronomy, Stony Brook University, Stony
Brook, NY 11794-3800, USA}

\author{P.~P.~Orth}
\affiliation{Ames Laboratory, Ames, Iowa 50011, USA}
\affiliation{Department of Physics and Astronomy, Iowa State University, Ames,
Iowa 50011, USA}

\author{M.~S.~Scheurer}
\affiliation{Institute for Theoretical Physics, University of Innsbruck, Innsbruck
A-6020, Austria}

\author{R.~Prozorov}
\email[Corresponding author: ]{prozorov@ameslab.gov}
\affiliation{Ames Laboratory, Ames, Iowa 50011, USA}
\affiliation{Department of Physics and Astronomy, Iowa State University, Ames,
Iowa 50011, USA}

\date{Submitted: 5 October 2021; Accepted: 26 February 2022}

\begin{abstract}
We study the evolution of temperature-dependent resistivity with
added point-like disorder induced by 2.5 MeV electron irradiation
in stoichiometric compositions of the ``3-4-13'' stannides, $(\text{Ca,Sr})_{3}(\text{Ir,Rh})_{4}\text{Sn}_{13}$.
Three of these cubic compounds exhibit a proposed microscopic coexistence of
charge-density wave (CDW) order and superconductivity (SC), while $\text{Ca}_{3}\text{Rh}_{4}\text{Sn}_{13}$
does not develop CDW order. As expected, the CDW transition temperature,
$T_{\text{CDW}}$, is universally suppressed by irradiation in all
three compositions. 
The superconducting transition temperature, $T_{c}$, behaves in a
more complex manner. In $\text{Sr}_{3}\text{Rh}_{4}\text{Sn}_{13}$,
it increases initially in a way consistent with a direct competition
of CDW and SC, but quickly saturates at higher irradiation doses. In the other three
compounds, $T_{c}$ is monotonically suppressed by irradiation. The
strongest suppression is found in $\text{Ca}_{3}\text{Rh}_{4}\text{Sn}_{13}$,
which does not have CDW order. We further examine this composition by measuring the London penetration depth, $\lambda(T)$, from which we derive the superfluid density. The result unambiguously points to a weak-coupling, full single gap, isotropic superconducting state. Therefore, we must explain two seemingly incompatible experimental observations: a single isotropic superconducting gap and a significant suppression of $T_{c}$ by non-magnetic disorder. We conduct a quantitative theoretical analysis based on a generalized Anderson theorem which points to an unconventional multiband $s^{+-}$-pairing state where the sign of the order parameter is different on one (or a small subset) of the smaller Fermi surface sheets, but remains isotropic and overall fully-gapped. 
\end{abstract}

\maketitle

\section{Introduction}

Extensive studies over the past few decades have identified a number
of characteristics that are common in unconventional superconductors.
First, unconventional superconductivity (SC) often develops in cooperation,
competition, or close proximity to other electronic long-range orders.
Second, non-Fermi-liquid behavior is often observed in the normal
state around the superconducting ``dome''. This behavior can be
associated with proximity to a putative quantum critical point (QCP)
inside the dome \cite{Mathur,Belitz2005,Monthoux,Keimer,Norman2011,Levchenko2013,CPLT,Khodas2020,Chubukov2020}. A QCP occurs when a continuous second-order phase transition is driven at $T=0$ by a non-thermal parameter, such as composition \cite{Hashimoto2012,Wang2018,Joshi2020},
pressure \cite{Mathur,Cheng2015,Park2011}, magnetic field \cite{Custers,Paglione,Budko}
or disorder \cite{Mutka,Belitz2005,Shibauchidisorder,NbSe2disorder}. It has
been suggested that the fluctuations of the coexisting order parameter
may act as a ``glue" for Cooper pairing of conduction electrons  \cite{Mathur,Norman2011,Chubukov2020,Cheng2015,Monthoux}. This approach
is actively discussed for high$-T_{c}$ cuprates \cite{Keimer,Abanov2003,Norman2011,CPLT},
heavy-fermion materials \cite{QCPhf,Park2011}, and it is particularly relevant
in iron based superconductors where there is a significant range of microscopic coexistence of antiferromagnetic and superconducting phases \cite{CB,DCJ2010,paglione2010,carrington,Shibauchireview,Eulers,JHChu,Chubukov2020,Khodas2020}.

In most known cases, the order parameter tuned to a QCP is spin-density wave (SDW). However, charge-density wave (CDW) ordering is another candidate if it can be continuously suppressed \cite{Goh2015,Klintberg2012,Veiga2020}. While pressure or magnetic field tuning is particularly useful for singular compositions, it is desirable to find superconducting systems tuneable through QCP by doping, allowing for a wider range of different types of measurements. Unfortunately, in most known CDW/SC systems, CDW ordering appears only in single compositions.

A CDW is formed when electronic energy is sufficiently lowered
by opening an energy gap on parts of the Fermi surface \cite{Peierls,CDW,Pouget2016}.
Usually this leads to the formation of a spatially-modulated charge-density state. In a one-dimensional case, a straightforward nesting determines the modulation wave-vector \cite{Peierls}, as observed in one-dimensional organic materials \cite{Pouget2016,Jerome}. In two dimensional systems such as
transition metal dichalcogenides, $\text{2H-NbSe}_{2}$ \cite{NbSe2nesting}, 2H-$\text{TaSe}_{2}$ \cite{TaSe2nesting} and 2H-$\text{TaS}_{2}$ \cite{TaS2nesting}), the nesting mechanism is not so obvious. It is even more complicated in three dimensions, such as our 3-4-13 cubic practically isotropic compounds.  

The charge density wave in Remeika 3-4-13 series \cite{Remeika} was studied by a variety of the techniques, and has a number of anomalous features. Modulation of the crystal lattice with the wavevector q = ($1/2$, $1/2$, 0) was found in $\text{Ca}_{3}\text{Ir}_{4}\text{Sn}_{13}$ \cite{Mazzone2015} and $\text{Sr}_{3}\text{Ir}_{4}\text{Sn}_{13}$ \cite{Tompsett2014}, which does not seem to correspond to nesting conditions. Similarly, in $(\text{Sr})_{3}(\text{Rh})_{4}\text{Sn}_{13}$ computational mode decomposition has revealed the same q-vector ($1/2$, $1/2$, 0) \cite{Goh2015}.  In a closely structurally related compound $\text{Yb}_{3}\text{Co}_{4}\text{Ge}_{13}$, charge density modulation was found to depend on sample stoichiometry \cite{Dalton2021}. The EXAFS phase derivative analysis supports the CDW-like formation by revealing different bond distances between two tin sites [Sn1(2)-Sn2] below and above $T_{CDW}$ in the (110) plane in $\text{Sr}_{3}\text{Ir}_{4}\text{Sn}_{13}$ \cite{Wang2017}. XANES spectra at the Ir L3-edge and Sn K-edge demonstrated an increase (decrease) in the unoccupied (occupied) density of Ir 5d-derived states and a nearly constant density of Sn 5p-derived states. A close relationship was suggested to exist between local electronic and atomic structures and the CDW-like phase in the $\text{Sr}_{3}\text{Ir}_{4}\text{Sn}_{13}$ single crystal \cite{Wang2017}.

Inelastic neutron scattering data point towards a displacive structural transition in the $\text{Ca}_{3}\text{Ir}_{4}\text{Sn}_{13}$ compound arising from the softening of a low-energy phonon mode with an energy gap of $\Delta=$120~K \cite{Mazzone2015}. Softening of the acoustic phonon modes was also suggested by ultrafast spectroscopy study in $\text{Sr}_{3}\text{Ir}_{4}\text{Sn}_{13}$ revealing also a correlation of optical phonons with the transition \cite{Luo2016}. 
Reduction of the magnetic susceptibility and a sign change of the Hall resistivity could be due to transformation of the Fermi surface below $T_{CDW}$ in $\text{Ca}_{3}\text{Ir}_{4}\text{Sn}_{13}$ and $\text{Sr}_{3}\text{Ir}_{4}\text{Sn}_{13}$ \cite{Wang2015}. This conclusion is supported optical reflection study \cite{mechanismCDW3413} and by the anomalies in the NMR Knight shift \cite{CNKuo2014}. Splitting of the NMR lines in the CDW phase imply local distortions of the Sn2 icosahedra  \cite{CNKuo2014}. On the other hand, the detailed structure of Remeika series compounds may be much more complicated than usually assumed primitive cubic cell \cite{ChemistryRareEarth,CrystEngComm,JPhysCM}. 

The influences of different structural models on the calculated electronic structures of some 3:4:13 compounds were discussed in Ref.~\onlinecite{Dalton2015}.
Furthermore, unconventional character of CDW and second order phase transition have been found by X-ray structural studies in a related to this work compositions,
$(\text{La,Ce})_{3}(\text{Ir,Rh})_{4}\text{Sn}_{13}$ \cite{Suyama2018}. Various mechanisms of CDW formation in these materials are suggested \cite{mechanisms}.
Importantly, the 3-4-13 compounds, specifically $(\text{Ca,Sr})_{3}(\text{Ir,Rh})_{4}\text{Sn}_{13}$ seem to exhibit a putative QCP under the dome of superconductivity \cite{Klintberg2012}. 

Here we study the influence of controlled disorder on CDW and superconductivity in 3-4-13 superconductors,
to uncover the connection between the two quantum orders and the nature
of the superconducting state. Intuitively, the opening of the CDW gap should decrease the density
of states at the Fermi surface and thus lower the superconducting
transition temperature in materials where CDW and superconductivity
coexist \cite{Gabovich}. This is indeed frequently observed \cite{CDWSCcompetition}.
In the $\text{YBa}_{2}\text{CuO}_{6-\delta}$, the CDW transition is enhanced
when superconductivity is suppressed by magnetic field, and the superconducting
transition temperature increases when CDW ordering is suppressed with
pressure \cite{YBCO}. In the transition metal dichalcogenides, $\text{2H-NbSe}_{2}$, 2H-$\text{TaS}_{2}$, and 2H-$\text{TaSe}_{2}$, 2.5 meV electron irradiation experiments suggested that long-range ordered CDW directly competes with SC so that superconducting transition temperature, $T_{c}$, increases with the introduction of disorder \cite{Mutka}. However, this simple competition between CDW and SC is only part of the story. Further irradiation experiments showed that as soon as the long-range CDW order breaks down above approximately   $6\times10^{18}$ electrons per $\text{cm}^{2}$, $T_{c}$ starts to decrease rapidly, initially in a step like fashion \cite{NbSe2disorder}. This implies that CDW also helps superconductivity which benefits from softening of the phonon modes due to long-range CDW order \cite{NbSe2disorder}. Phonon softening near the $T_{CDW}$ transition is also observed in $\text{Sr}_{3}\text{Ir}_{4}\text{Sn}_{13}$ \cite{Luo2016}, $\text{Sr}_{3}\text{Rh}_{4}\text{Sn}_{13}$ \cite{Cheung2018}, and $\text{Ca}_{3}\text{Ir}_{4}\text{Sn}_{13}$ \cite{Mazzone2015}. 
Furthermore, later studies of $\text{2H-NbSe}_{2}$ showed that in systems
with electron-phonon pairing mechanism, the largest superconducting gaps occur
in the regions of the Fermi surface connected by the CDW nesting vector.
\cite{Kiss2007}

\begin{figure}[tb]
\centering
\includegraphics[width=0.95\linewidth]{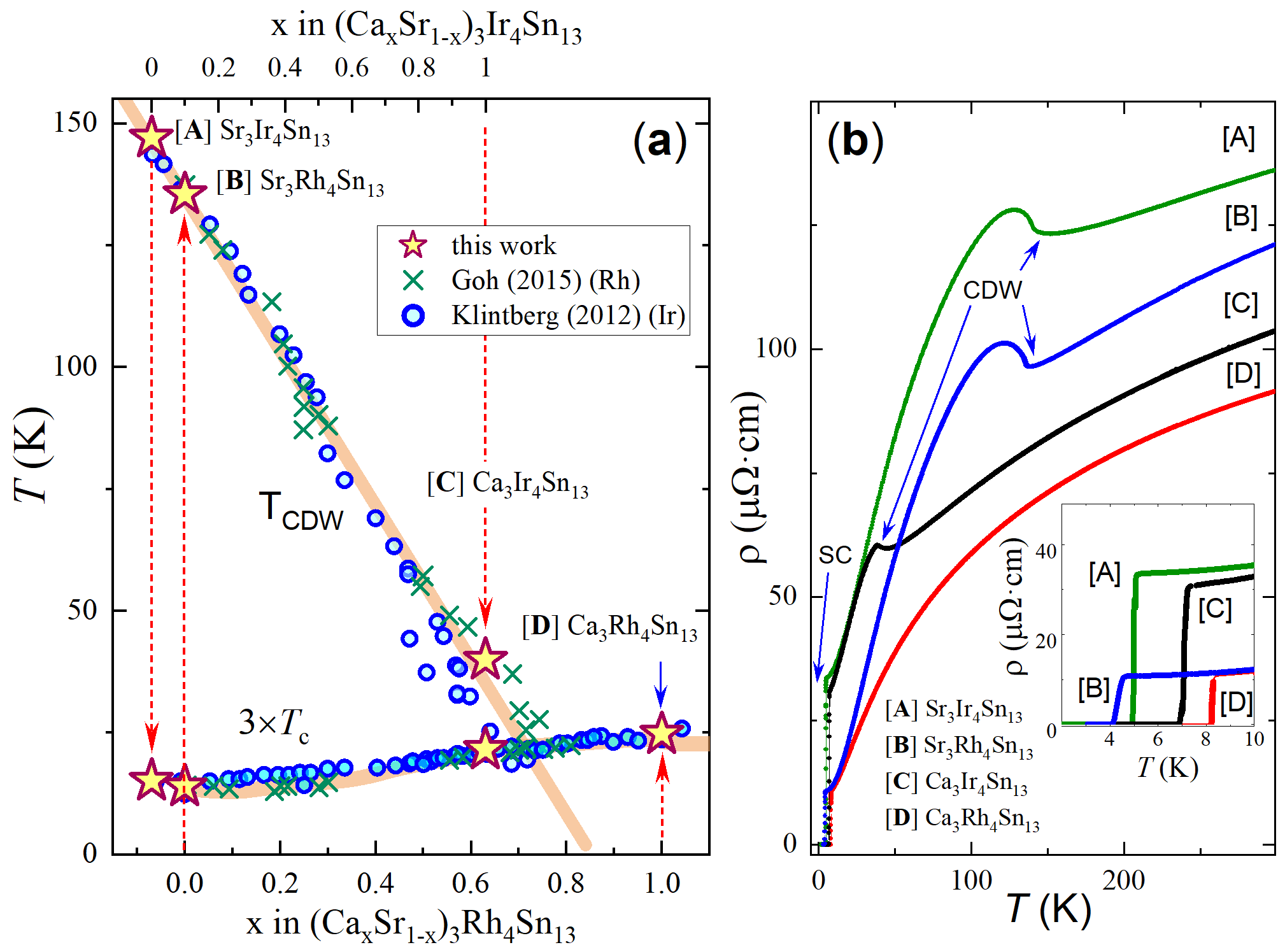} 
\caption{(\textbf{a}) Combined phase diagram of 3-4-13 compounds as determined
from measurement (Ca$_{x}$Sr$_{1-x}$)$_{3}$Rh$_{4}$Sn$_{13}$
(bottom axis, cross symbols)~\cite{Klintberg2012} and (Ca$_{x}$Sr$_{1-x}$)$_{3}$Ir$_{4}$Sn$_{1}3$
(top axis, square symbols)~\cite{Goh2015}. The phase diagram for
(Ca$_{x}$Sr$_{1-x}$)$_{3}$(Rh, Ir)$_{4}$Sn$_{13}$ was mapped
using a combination of doping and pressure. The positions of the samples
used in this study are shown by yellow-red stars. (\textbf{b}) Temperature-dependent
resistivity of (Sr, Ca)$_{3}$(Rh, Ir)$_{4}$Sn$_{13}$ samples selected
for electron irradiation in this study. The inset zooms at the superconducting
transition.}
\label{fig:PhaseDiagram} 
\end{figure}

The 3-4-13 family of compounds is well-suited for studying the relationship
between CDW and superconductivity. Their CDW transition can be tuned
through a broad range of temperatures by the selection of different
elements or by the application of pressure. The suppression of CDW
ordering extrapolates to a region where the resistivity exhibits non-Fermi
liquid behavior, suggesting the existence of a QCP in the phase diagram.
This QCP was first discovered in $(\text{Ca}_{x}\text{Sr}_{1-x})_{3}\text{Ir}_{4}\text{Sn}_{13}$ 
compounds at the pressure of about 20 kbar \cite{Goh2015}, and was
later found to be accessible via doping in the $(\text{Ca}_{x}\text{Sr}_{1-x})_{3}\text{Rh}_{4}\text{Sn}_{13}$
series at around $x=0.9$ \cite{Klintberg2012,Cheung2018}. The structural nature
of the QCP was confirmed using x-ray diffraction, showing the continuation
of the CDW ordering inside of the superconducting dome \cite{Veiga2020}.
The summary phase diagram as determined from these measurements, with the location of our samples marked, is shown in Fig.~\ref{fig:PhaseDiagram}(a). 

Experimentally, it is determined that CDW materials exhibit mostly
conventional electron-phonon mechanism of superconductivity \cite{CDWphonon}. In the 3-4-13 compounds most studies, including this work, are consistent with a single isotropic gap weak-coupling superconductivity. Thermal conductivity measurements of  $\text{Ca}_{3}\text{Ir}_{4}\text{Sn}_{13}$ found a vanishing residual linear
term and a weak increase with applied magnetic field, consistent with a full gap with small or no anisotropy \cite{Thermalconductivity}. Heat capacity measurements show exponential decrease at low temperatures \cite{Kase2011} and a linear magnetic field dependence \cite{Wang2012}, which also agree with a full-gap superconducting state. Temperature dependence of the London penetration depth, $\lambda(T)$, determined from lower critical field measurements \cite{Wang2015} as well as this work discussed later in the text provide strong evidence of a fully-gaped superconducting state. Even more so we found a perfect agreement of the data with $\lambda(T)$, expected from the weak-coupling isotropic BCS theory, parameter-free, both close to $T\rightarrow0$ and in the full temperature range.

On the other hand, an apparent enhancement of the electronic specific-heat jump at $T_c$ in  $(\text{Sr})_{3}(\text{Ir})_{4}\text{Sn}_{13}$ and $(\text{Sr})_{3}(\text{Rh})_{4}\text{Sn}_{13}$ was interpreted as a sign of a strong-coupling nature of superconductivity in these compounds \cite{Lue2016}. 
Furthermore, there are signs of strong coupling superconductivity in heat capacity measurements around the QCP region \cite{Goh2,Luo2018,swave,Akimitsu}, which could also be due to the contribution of critical quantum fluctuations. Muon-spin rotation ($\mu$SR) experiments under pressure find that the superfluid density strongly
increases when the system is tuned closer to the QCP in $\text{Ca}_{3}\text{Ir}_{4}\text{Sn}_{13}$ \cite{swave}. While $\mu$SR measurements of both $\text{Ca}_{3}\text{Ir}_{4}\text{Sn}_{13}$ \cite{Biswas2014} and $\text{Sr}_{3}\text{Ir}_{4}\text{Sn}_{13}$ \cite{multibandmusR} agree with a single isotropic gap, they also could not rule out possible two-gap superconductivity with two very different gaps on different Fermi surface sheets. The same group discusses possible multi-band physics from the nuclear magnetic resonance (NMR) measurements \cite{multibandNMR}. We note that in 2D CDW/SC $\text{2H-NbSe}_{2}$, angle-resolved photoemission spectroscopy (ARPES) \cite{Science}, specific heat \cite{Boaknin} and London penetration depth \cite{Prozorov} measurements found strong evidence for multi-gap superconductivity. In $\text{Sr}_{3}\text{Ir}_{4}\text{Sn}_{13}$, possible importance of multi-band effects was identified in electronic band-structure study where at least four sheets of the Fermi surface  with sizes differing by a factor of nearly 20 were found \cite{FS}. 

Regarding the superconducting gap(s) anisotropy, most measurements are consistent with a fully-gapped isotropic superconducting state described by a weak-coupling Bardeen-Cooper-Schrieffer (BCS) theory \cite{BCS}, which is natural for a phonon-mediated attractive pairing potential. In the case of SDW antiferromagnetic fluctuations as in the cuprates, a sign-changing $d-$wave pairing is favored \cite{Annett1996}. In the present case of CDW/SC compounds, the pairing type is an open question and our present work strongly suggests a possibility of an unconventional multiband $s^{+-}$-pairing state where the sign of the order parameter is different on one (or a small subset) of the smaller Fermi surface sheets, but remains isotropic and overall fully-gapped. Such a state will manifest itself only in select experiments, such as the response to a non-magnetic disorder. 
On a general note, there is currently significant revived interest in superconductivity in seemingly conventional compounds, such as elemental niobium where the response to disorder has helped to reveal anisotropic strong-coupling superconductivity \cite{Sauls2022}, or in the case of a Dirac semi-metal from our earlier work \cite{TimmonsPdTe2}. 

It should be noted that thermodynamic measurements are not sensitive to the sign of the order parameter. On the other hand, studying the variation of $T_{c}$ when changing the non spin-flip (non magnetic) scattering rate is a phase-sensitive method that provides insights into the nature of the order parameter and pairing mechanisms \cite{CaK1144,TimmonsPdTe2}. In the well-known
limit of an isotropic single-band $s-$wave superconductor, $T_{c}$ is not affected by weak non-magnetic disorder, known as the ``Anderson theorem" \cite{Anderson,AG}. In a stark contrast, the transition temperature in materials
with anisotropic gap(s)~\cite{Hohenberg1964, GolubovMarzin,Sauls2022}, or sign-changing $d-$wave superconductivity in the cuprates \cite{tccuprates}, as well as $s^{+-}$ pairing states in iron-based
superconductors \cite{IBSTcsuppression}, is strongly affected by
non-magnetic disorder. A generalized treatment extending the original Abrikosov-Gor'kov theory \cite{AG} for anisotropic order parameters is given by Openov \cite{Openov1997,Openov2004}, and it can be easily extended to a multiband case with different gap amplitudes \cite{Cho2022}. In multi-band and multi-orbital
systems, particularly in the presence of spin-orbit coupling, the
suppression of $T_{c}$ is expected to be somewhere in between these
two limits \cite{TimmonsPdTe2,DisorderSOCFu,OurDisorderSOC,BrydonScattering}. Importantly, combined with independent measurements of the superfluid density and theoretical calculations that take into account particular crystal and electronic structure, the evolution of $T_{c}$ with disorder is a powerful tool to extract important information about the superconducting order parameter \cite{TimmonsPdTe2,Sauls2022}.

In this work, we use artificial point-like disorder to study the relationship between superconductivity and CDW ordering in the stoichiometric compounds of the $(\text{Ca,Sr})_{3}(\text{Ir,Rh})_{4}\text{Sn}_{13}$ ``3-4-13" Remeika series. The low-temperature (20 K) 2.5 MeV electron
irradiation produces vacancy-interstitials ``Frenkel pairs'', which
leave a metastable population of vacancies upon warming up to room temperature
due to very different rates of diffusion of vacancies and interstitials
\cite{Damask1963,Thompson1969,Westerveld1989}. This leads to a residual resistivity
increase which is monotonic with the irradiation dose, reflecting
the increase in the scattering rate. We find that in 3-4-13 compounds, the CDW transition is universally suppressed by disorder. We also observe a weak increase of the superconducting transition temperature,
$T_{c}$, in $\text{Sr}_{3}\text{Rh}_{4}\text{Sn}_{13}$, and a non-linear scattering-rate
dependence of $T_{c}$ in $\text{Sr}_{3}\text{Ir}_{4}\text{Sn}_{13}$ and $\text{Ca}_{3}\text{Ir}_{4}\text{Sn}_{13}$.
Contrary to the expectations for conventional superconductivity, $T_{c}$ is rapidly suppressed with disorder in $\text{Ca}_{3}\text{Rh}_{4}\text{Sn}_{13}$,
which does not exhibit any long-range CDW order. This behavior became puzzling when precision London penetration depth measurements found a full and isotropic single superconducting gap in this compound.  This apparent contradiction is resolved by a detailed theoretical analysis of possible pairing states, which provides strong argument in favor of unconventional multiband $s^{+-}$-pairing state where the sign of the order parameter is different on one (or a small subset) of the smaller Fermi surface sheets, but remains overall fully-gapped. We note that the influence of atomic defects produced by rapid quenching from high temperatures in $\text{Ca}_{3}\text{Rh}_{4}\text{Sn}_{13}$ was studied thirty years ago using x-ray spectroscopy \cite{Westerveld1989}. The observed reduction of $T_c$~\cite{Westerveld1987} was attributed to the creation of Sn-Ca ions exchange anti-sites. Unfortunately, no physical properties, for example conductivity, were measured, hence the dimensionless scattering rate was not determined. The authors of Ref.~\cite{Westerveld1989} speculated that $T_c$ decreased due to the suppression of the density of states at the Fermi level due to the disturbance of the Ca-Ca bond length. However, we believe that it is more likely that they dealt with the same unconventional mechanism as proposed in our report here. As discussed below, our electron irradiation creates roughly one atomic defect per thousand formula units, which has no appreciable effect on the density of states.

The paper is organized as follows: details of sample preparation and methods are provided in Section~\ref{ExpDetails}. The experimental results for all compounds can be found in Section~\ref{Sec:Results} and theoretical analysis in Section~\ref{Sec:Discussion}. Finally, Section~\ref{Sec:Conclusion}
summarizes our findings.

\section{Experimental methods\label{ExpDetails}}

Single crystals of (Ca,Sr)$_{3}$(Rh,Ir)$_{4}$Sn$_{13}$ were grown
using a high temperature self-flux method, as described in Ref. \cite{Wang2012}.
X-ray diffraction (XRD) data were taken with Cu K$_{\alpha}$ ($\lambda =$ 0.15418 nm) radiation of a Rigaku Miniflex powder diffractometer, and the elemental analysis was performed using an energy-dispersive x-ray spectroscopy (EDX) in a JEOL JSM-6500 scanning electron microscope.

Electrical resistivity measurements were conducted in a \textit{Quantum Design} PPMS using a conventional four-probe method. The contacts to the crystal surface were made by soldering silver wires with tin \cite{anisotropy,SUST}. The contact resistance is below 100 $\mu\Omega$,
and they are sufficiently mechanically stable to withstand electron
irradiation \cite{TimmonsRSI}. The samples for resistivity measurements
were cut and polished from single crystals, with typical sample sizes of (1-2)$\times$0.3$\times$0.1 mm$^{3}$. The long sample axis was arbitrary with respect to the cubic structure of these crystals. Standard resistivity runs were made on both cooling and heating, with negligible hysteresis. 

The variation of the in-plane London penetration depth, $\Delta\lambda(T)$,
was measured using a sensitive self-oscillating tunnel-diode resonator
(TDR) described in detail elsewhere \cite{VanDegrift1975RSI,Prozorov2000PRB,Prozorov2000APL,Prozorov2021}.
In brief, the TDR circuit resonates at approximately 14 MHz, and the
frequency shift, which is proportional to the sample magnetic susceptibility,
is measured with precision better than one part per billion (ppb).
The coefficient of proportionality that includes the demagnetization
correction is measured directly by pulling the sample out of the resonator at base temperature \cite{Prozorov2021}. This technique was developed specifically to detect minute changes in the London penetration depth
and is now considered one of the sensitive tools for studying the
anisotropy of the superconducting order parameter \cite{Prozorov2006SST,Prozorov2011RPP_review,Giannetta2021}.
We use this technique to determine the superconducting gap structure, as well as to show that we do not induce magnetic states with disorder, and that our crystals are very homogeneous.

Point-like disorder was introduced at the SIRIUS facility in the Laboratoire
des Solides Irradi\'{e}s at \'{E}cole Polytechnique, Palaiseau, France. Electrons accelerated in a pelletron-type accelerator to 2.5 MeV knock out ions creating vacancy - interstitial Frenkel pairs \cite{Damask1963,Thompson1969}. During irradiation the sample is held in liquid hydrogen at around 20 K. The low-temperature environment is needed not only to remove the significant amount of heat produced by sub-relativistic electrons upon collisions, but also to prevent the immediate recombination and migration of produced atomic defects. The acquired irradiation dose
is determined by measuring the total charge collected by a Faraday
cage located behind the sample. As such, the acquired dose is measured in the ``natural" units of $\text{C/cm}^{2}$, 
which is equal to $1\:\textrm{C}\equiv1/e\approx6.24\times10^{18}$
electrons per cm$^{2}$. Upon warming the sample to room temperature, the interstitials, which have a lower barrier of diffusion \cite{Damask1963,Thompson1969}, migrate to various sinks (dislocations, surfaces etc). This leaves
a metastable population of vacancies. The resultant vacancy density is determined by the highest temperature the sample was exposed to. In most materials, including 3-4-13, vacancies are stable as verified by the transport measurements of the same samples years apart and even if the density would slowly change, the resistivity measurement provides a snapshot of the current scattering rate in a particular sample. This is the point-like disorder discussed in this paper \cite{PRX,BaKMathesson}. Practically, the
level of disorder induced by the irradiation is gauged experimentally by the change of resistivity well above the CDW transition, at the room temperature, where the carrier density is roughly constant across all compositions and the only change in resistivity comes from the difference in the residual resistivity. We also calculated the number of defects per formula unit (dpf) numerically using specialized ``SECTE" software developed in \'{E}cole Polytechnique (Palaiseau, France) specifically
to describe ion-resolved knock-out cross-sections for MeV-range electron irradiation. The summary of the results for our four compositions is given in Table \ref{tab:dpf}. The first three columns show partial cross-sections of the defects created
upon head-on collision of a 2.5 MeV electron with an indicated ion,
assuming the same value of the barrier for ion displacement from its position, $E_{d}=25\:\textrm{eV}$. This is a ``generic'' number
for intermetallic compounds, usually in the range of tens of eV, and it can be calculated using methods of molecular dynamics \cite{Damask1963,Thompson1969}.
However, its exact value is not very important for our rough estimates. The fourth column shows the total cross-section of knocking out any ion by using molecular weight averaging of the partial cross-sections. The last column shows the estimated number of defects per formula unit ignoring possible annealing upon warming up after irradiation at 20 K. The realistic percentage lost in that process varies from almost no annealing to about $10\:\%-30\:\%$, for example measured by in-situ resistivity
in iron pnictides \cite{BaKMathesson}. Our SECTE calculations show
that electron irradiation of the 3-4-13 compounds creates less than 1 defect of any kind per 1000 formula units, which cannot alter the
chemical or electronic nature of the material. This also means that
the defects are well-separated and can be treated as point-like in
the dilute limit. This disorder is much ``softer" than that induced by rapid quenching from high temperatures used in earlier experiments \cite{Westerveld1989}.
Importantly, electron irradiation does not ``dope" the system as was shown directly by Hall resistivity measurements \cite{BaKMathesson}. In the present case, even if there was some induced variation of stoichiometry, $T_c(x)$ of 3-4-13 compounds is practically flat and could not result in the systematic shift observed. We note that chemical inhomogeneity and disorder may lead to the significant spread of $T_c$ \cite{Slebarski2016}. This, however, would change the observed superfluid density from exponential to a power law at low temperatures. 

\begin{table}[tb]
\caption{\label{tab:dpf}Head-on knock out partial cross-sections by 2.5 MeV
electron irradiation ($1\:\textrm{barn}=1\times10^{-24}\:\text{cm}^{2}$).
The last column shows the number of defects created per formula unit,
per 1 $\text{C}/\text{cm}^{2}$. Roughly 1 defect per 1000 formula
units is created. This is sufficiently close to the dilute limit to avoid significant compositional
or electronic change.}

\begin{ruledtabular}
\begin{tabular}{cccccc}
\multirow{2}{*}{compound} & Sr/Ca & Ir/Rh & Sn & total $\sigma$ & dpf$\times10^{-3}$\tabularnewline
\cline{2-6} \cline{3-6} \cline{4-6} \cline{5-6} \cline{6-6} 
 & barn & barn & barn & barn & per 1 $\text{C}/\text{cm}^{2}$\tabularnewline
\hline 
$\text{Sr}_{3}\text{Ir}_{4}\text{Sn}_{13}$ & 139 & 261 & 148 & 181 & 1.13\tabularnewline
$\text{Sr}_{3}\text{Rh}_{4}\text{Sn}_{13}$ & 139 & 158 & 145 & 147 & 0.92\tabularnewline
$\text{Ca}_{3}\text{Ir}_{4}\text{Sn}_{13}$ & 79 & 258 & 145 & 177 & 1.11\tabularnewline
$\text{Ca}_{3}\text{Rh}_{4}\text{Sn}_{13}$ & 84 & 155 & 143 & 142 & 0.89\tabularnewline
\end{tabular}
\end{ruledtabular}
\end{table}

\begin{figure}[tb]
\centering
\includegraphics[width=0.85\linewidth]{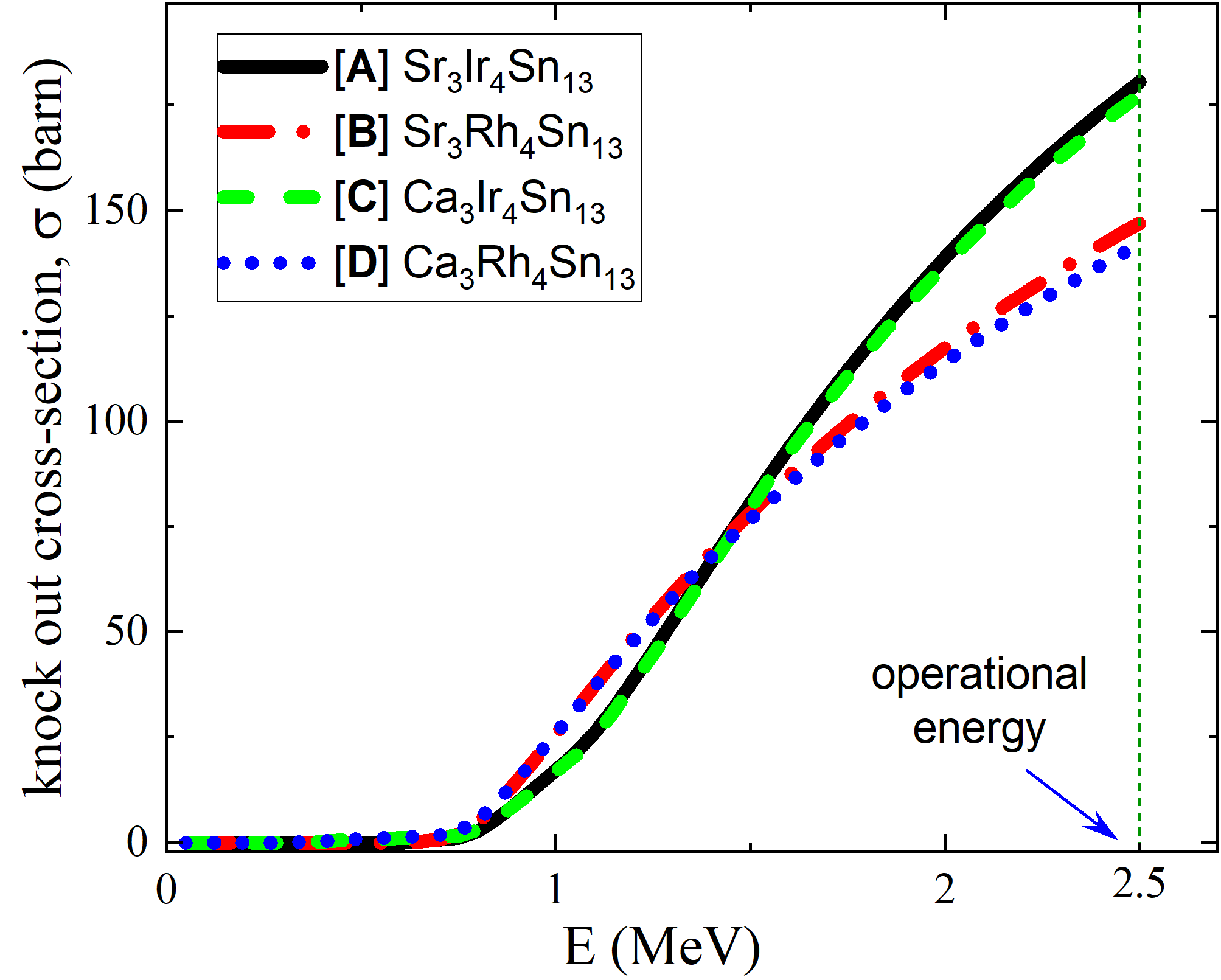}
\caption{Total knock-out cross-sections for studied compounds as function of
electron energy. Our operating energy of 2.5 MeV is marked by a dotted
line. The difference between Ca/Sr pairs is negligible and is not
large between Ir/Rh, being about 30 barn larger for Ir compounds.}
\label{fig:Cross-Section}
\end{figure}

The comparison of the total cross-sections as function of electron
energy for the studied compounds is shown in Fig.~\ref{fig:Cross-Section}.
There is practically negligible differences between Ca-(Ir/Rh)-Sn and
Sr-(Ir/Rh)-Sn compounds and quite small differences between (Ca/Sr)-Ir-Sn
and (Ca/Sr)-Rh-Sn, where in Ir compounds the cross-sections are larger
by about 30 barn. The resulting numbers at our operating frequency
of 2.5 MeV are summarized in Table \ref{tab:dpf}.

In our experiments, the same physical samples were measured before
and after electron irradiation, thus avoiding uncertainties from possible
variation of stoichiometry within each batch, geometric factors and
other parameters unique to each sample. For most compositions, measurements
were performed on at least three samples to obtain as objective results
as possible, see Table \ref{tab:Normal-state-resistivity}.

\section{Experimental Results\label{Sec:Results}}

We now discuss the experimental results obtained in our irradiation
studies for the following compounds, ordered by a decreasing value
of $T_{\text{CDW}}$: $\text{Sr}_{3}\text{Ir}_{4}\text{Sn}_{13}$ {[}\textbf{A}{]},
$\text{Sr}_{3}\text{Rh}_{4}\text{Sn}_{13}$ {[}\textbf{B}{]}, $\text{Ca}_{3}\text{Ir}_{4}\text{Sn}_{13}$,
{[}\textbf{C}{]} and $\text{Ca}_{3}\text{Rh}_{4}\text{Sn}_{13}$ {[}\textbf{D}{]}.
This way, we are moving from left to right towards and beyond the
quantum critical point in the generic phase diagram shown in Fig.~\ref{fig:PhaseDiagram}(a).
The trend in the superconducting transition temperature, $T_{c}$,
is non-monotonic in this sequence, with $\text{Sr}_{3}\text{Rh}_{4}\text{Sn}_{13}$
having the lowest transition at $T_{c}=4.2$ K, and overall representing
a typical for unconventional superconductors shallow ``dome'' of
superconductivity. The characteristic transition temperatures and
resistivity values at room temperature in the pristine state were
determined by averaging the measurements of multiple samples as summarized
in Table \ref{tab:Normal-state-resistivity}:

\begin{table}[tb]
\caption{\label{tab:Normal-state-resistivity}Parameters of studied compositions
in the pristine state, including CDW and superconducting transition
temperatures, and the resistivity at room temperature averaged over
indicated number of samples, $N$.}

\begin{ruledtabular}
\begin{tabular}{lllll}
compound & $T_{\text{CDW}}\:(\mathrm{K})$ & $T_{c}$ (K) & $\rho_{\textrm{RT}}$~$(\mu\Omega\text{cm})$ & \multirow{1}{*}{$N$}
\tabularnewline
\hline 
$\text{Sr}_{3}\text{Ir}_{4}\text{Sn}_{13}$  & $145.2 \pm0.5$ & $5.11\pm0.03$ & $168\pm29$ & 3\tabularnewline
$\text{Sr}_{3}\text{Rh}_{4}\text{Sn}_{13}$  & $135.76\pm0.14$ & $4.59\pm0.1$ & $129\pm17$ & 9\tabularnewline
$\text{Ca}_{3}\text{Ir}_{4}\text{Sn}_{13}$  & $39.0\pm0.59$ & $7.17\pm0.02$ & $120\pm16$ & 3\tabularnewline
$\text{Ca}_{3}\text{Rh}_{4}\text{Sn}_{13}$  & no CDW & $8.29\pm0.01$ & $112\pm3.87$ & 3\tabularnewline
\end{tabular}
\end{ruledtabular}
\end{table}

The overall resistivity decreases with decreasing $T_{\text{CDW}}$,
which is particularly obvious from the measurements on the samples
selected for electron irradiation, shown in Fig.~\ref{fig:PhaseDiagram}(b).
That comparison also reveals similar slopes of the temperature-dependent
resistivity near room temperature. In the full temperature range,
the temperature dependence of the resistivity, $\rho\left(T\right)$,
is quite unusual. In all compounds, the resistivity in the metallic
phase, above $T_{\text{CDW}}$, extrapolates to a very high residual
resistivity, similarly to the tantalum dichalcogenides \cite{Cedomirnpj}.
The resistivity ``bump'' when crossing into the CDW phase (therefore,
lowering carrier density, hence increasing $\rho$) barely reaches
$10\:\%$ of the resistivity value at $T_{\text{CDW}}$, and a significant
decrease in the resistivity is observed on further cooling down to
low temperatures. This behavior suggests that the loss of the carrier
density due to the opening of the CDW gap is small, as would naturally
be expected for a three-dimensional CDW material. This is in line
with NMR measurements of $\text{Sr}_{3}\text{Rh}_{4}\text{Sn}_{13}$, which found
that only approximately $13\:\%$ of the total density of states is
lost in the CDW transition \cite{NMR2015}. The very high values of
$\rho$ obtained by linear extrapolation from high temperatures to
$T=0$, and the quick loss of resistivity upon CDW ordering suggest
significant contribution of charge-disorder scattering, similar to
that suggested by Naito and Tanaka for the transition metal dichalcogenides
\cite{Naito1982,Naito2nd}. Interestingly, a similar type of $\rho(T)$
behavior is observed in $\text{Ca}_{3}\text{Rh}_{4}\text{Sn}_{13}$, in which long-range
CDW is not observed. This may be indicate that despite the total suppression
of the long-range CDW ordering in that compound, short-range correlations
may persist similarly to the case of CDW suppression by disorder in
$\text{2H-NbSe}_{2}$ \cite{NbSe2disorder} and in doped ZrTe$_3$ \cite{degiorgi}.

\subsection{$\text{Sr}_{3}\text{Ir}_{4}\text{Sn}_{13}$}

\begin{figure}[tb]
\centering
\includegraphics[width=0.9\linewidth]{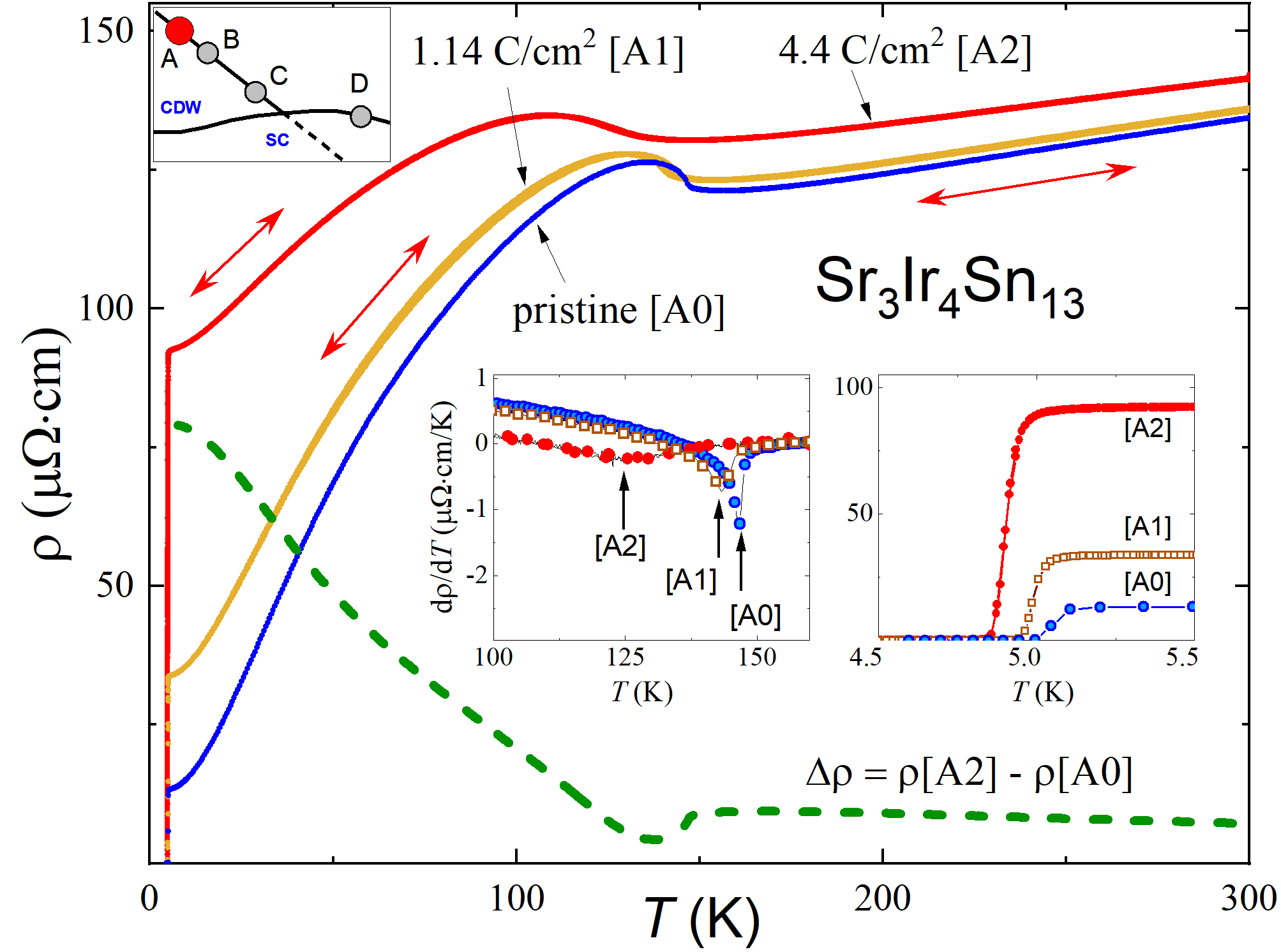} 
\caption{The evolution of temperature-dependent resistivity of $\text{Sr}_{3}\text{Ir}_{4}\text{Sn}_{13}$ in pristine (blue line) and after electron irradiation of 1.14 C/cm$^{2}$
(yellow), and 4.4 C/cm$^{2}$ (red). The green dashed line shows the
resistivity difference between the pristine and 4.4 C/cm$^{2}$ curves,
finding the Matthiessen's rule to be valid above $T_{\text{CDW}}$,
but, expectantly, grossly violated below. The small cartoon in the
top left corner indicates the sample's position on the generic phase
diagram. The left inset shows the resistivity derivative $d\rho/dT$
in the vicinity of the CDW transition, where the arrows show the positions
of the sharp features used to determine $T_{\text{CDW}}$. The right
inset zooms into the region around the superconducting transition. }
\label{fig:Sr3Ir4Sn13} 
\end{figure}

The temperature dependent electrical resistivity for $\text{Sr}_{3}\text{Ir}_{4}\text{Sn}_{13}$
and its evolution with electron irradiation are shown in the main
panel of Figure \ref{fig:Sr3Ir4Sn13}. The resistivity value for the
selected sample of $\text{Sr}_{3}\text{Ir}_{4}\text{Sn}_{13}$ is
in reasonable agreement with previous reports of 120 $\mu\Omega\cdot$
cm \cite{Luo2018,Wang2012}. Irradiation shifts the $\rho(T)$ curves
upward at high temperatures, but they remain nearly parallel to each
other above $T_{\text{CDW}}$. This can be seen in the plot of the
difference between the two curves $\Delta\rho=\rho(4.4\text{C}/\text{cm}^{2})-\rho(0\text{C}/\text{cm}^{2})$,
which is shown as the green line in Fig.~\ref{fig:Sr3Ir4Sn13}. Matthiessen's
rule is largely obeyed above the transition temperature, suggesting
that we are in a normal metallic state, albeit one with very high
residual resistivity. The minimum in the difference plot is caused
by the shift in the CDW transition temperature as irradiation disrupts
the long-range order. The suppression of that transition temperature
is shown in the left inset via a plot of the derivative of the resistivity,
$d\rho/dT$, with arrows indicating the location of $T_{\text{CDW}}$.
The superconducting transition temperature is monotonically suppressed
with disorder, and sharpens after irradiation.

\subsection{$\text{Sr}_{3}\text{Rh}_{4}\text{Sn}_{13}$}

\begin{figure}[tb]
\includegraphics[width=0.9\linewidth]{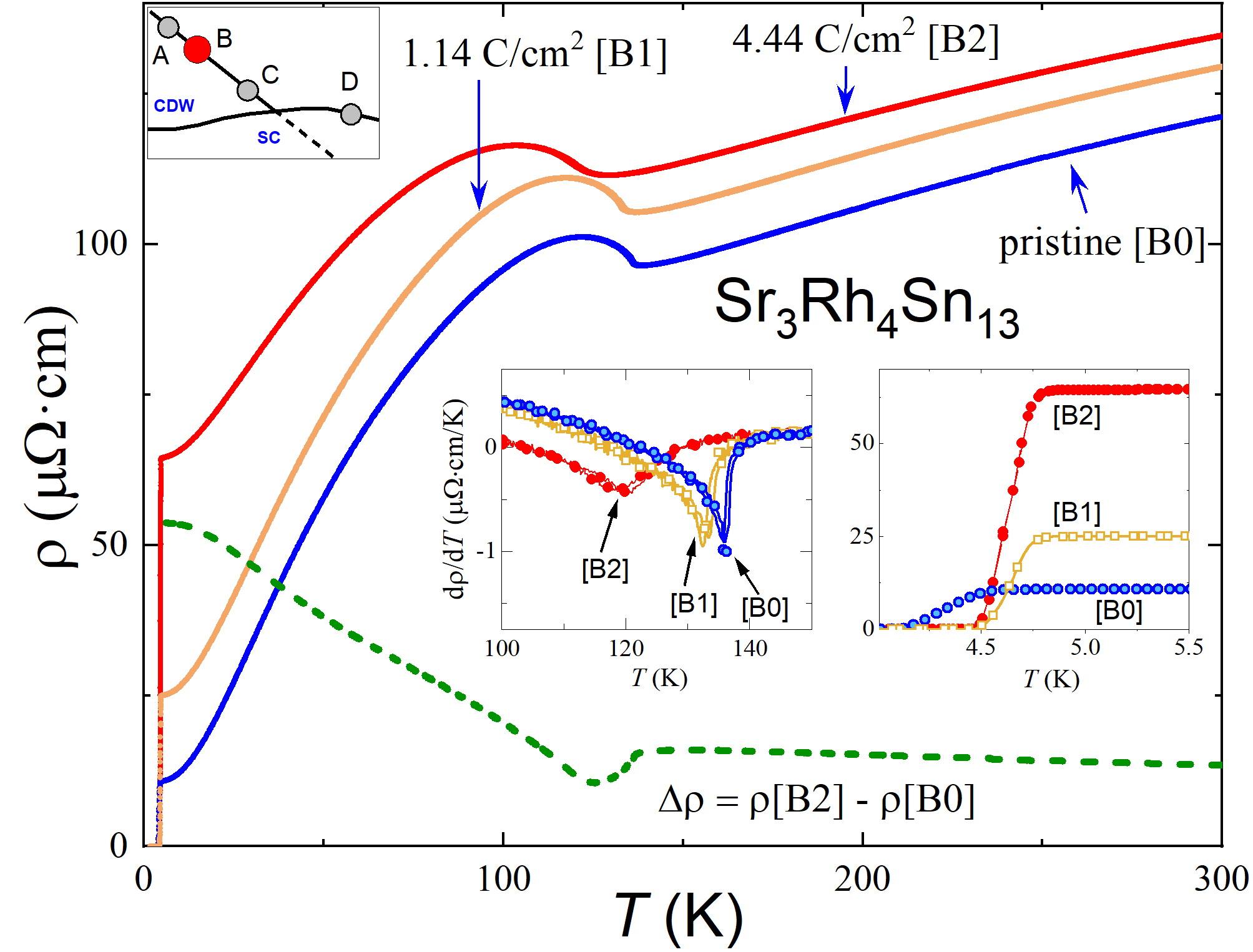} 
\centering
\caption{Temperature-dependent resistivity of $\text{Sr}_{3}\text{Rh}_{4}\text{Sn}_{13}$
before irradiation (blue curve), after 1.14 C/cm$^{2}$ irradiation
(yellow), and after 4.44 C/cm$^{2}$ (red) irradiations. The green
dashed line shows the resistivity difference between the pristine
and 4.44 C/cm$^{2}$ curves, finding the Matthiessen rule valid above
$T_{\text{CDW}}$, but violated below. The small cartoon in the top
left corner indicates the sample position on generic phase diagram.
The left inset shows the derivative of the resistivity in a region
around the CDW transition with arrows pointing to $T_{\text{CDW}}$.
This emphasizes the transition shift between the pristine state and
after 4.44 C/cm$^{2}$ dose of irradiation. The right inset zooms
into a region around the superconducting transition showing a non-monotonic
behavior of $T_{c}$: a small initial $T_{c}$ increase after 1.14
C/cm$^{2}$ irradiation, but only minimal changes in the behavior
between 1.14 and 4.44 C/cm$^{2}$. }
\label{fig:Sr3Rh4Sn13} 
\end{figure}

In $\text{Sr}_{3}\text{Rh}_{4}\text{Sn}_{13}$, similarly to $\text{Sr}_{3}\text{Ir}_{4}\text{Sn}_{13}$,
the CDW transition is monotonically suppressed with the increase of
disorder. However, $\text{Sr}_{3}\text{Rh}_{4}\text{Sn}_{13}$ is
the only compound in which the expected increase of the superconducting
transition temperature, $T_{c}$, with the suppression of CDW is actually
observed. The response of $T_{c}$ to disorder is distinctly non-linear,
with a significant initial increase which becomes smaller at higher
doses. Also, we found a larger variation of $T_{c}$ between the samples
from the same batch when performed initial screening, suggesting that
the superconducting state is sensitive to disorder either directly
or via the disruption of CDW order. It is possible that in $\text{Sr}_{3}\text{Ir}_{4}\text{Sn}_{13}$
the incipient superconductivity is too weak and $T_{c}\left(x\right)$
is too shallow to show any response to the suppressed CDW. In other
words, CDW is too strong. Then next in line, $\text{Sr}_{3}\text{Rh}_{4}\text{Sn}_{13}$,
has just right ratio of CDW and SC phases strength to see the effect. Of course, the $T_{c}$ is always monotonically
suppressed if CDW is not considered. The full range of resistivity
is shown for a selected representative sample in Fig.~\ref{fig:Sr3Rh4Sn13}(b).
Mattheissen's rule is largely obeyed above the CDW transition, similar
to $\text{Sr}_{3}\text{Ir}_{4}\text{Sn}_{13}$.

\subsection{$\text{Ca}_{3}\text{Ir}_{4}\text{Sn}_{13}$}

\begin{figure}[tb]
\centering
\includegraphics[width=0.9\linewidth]{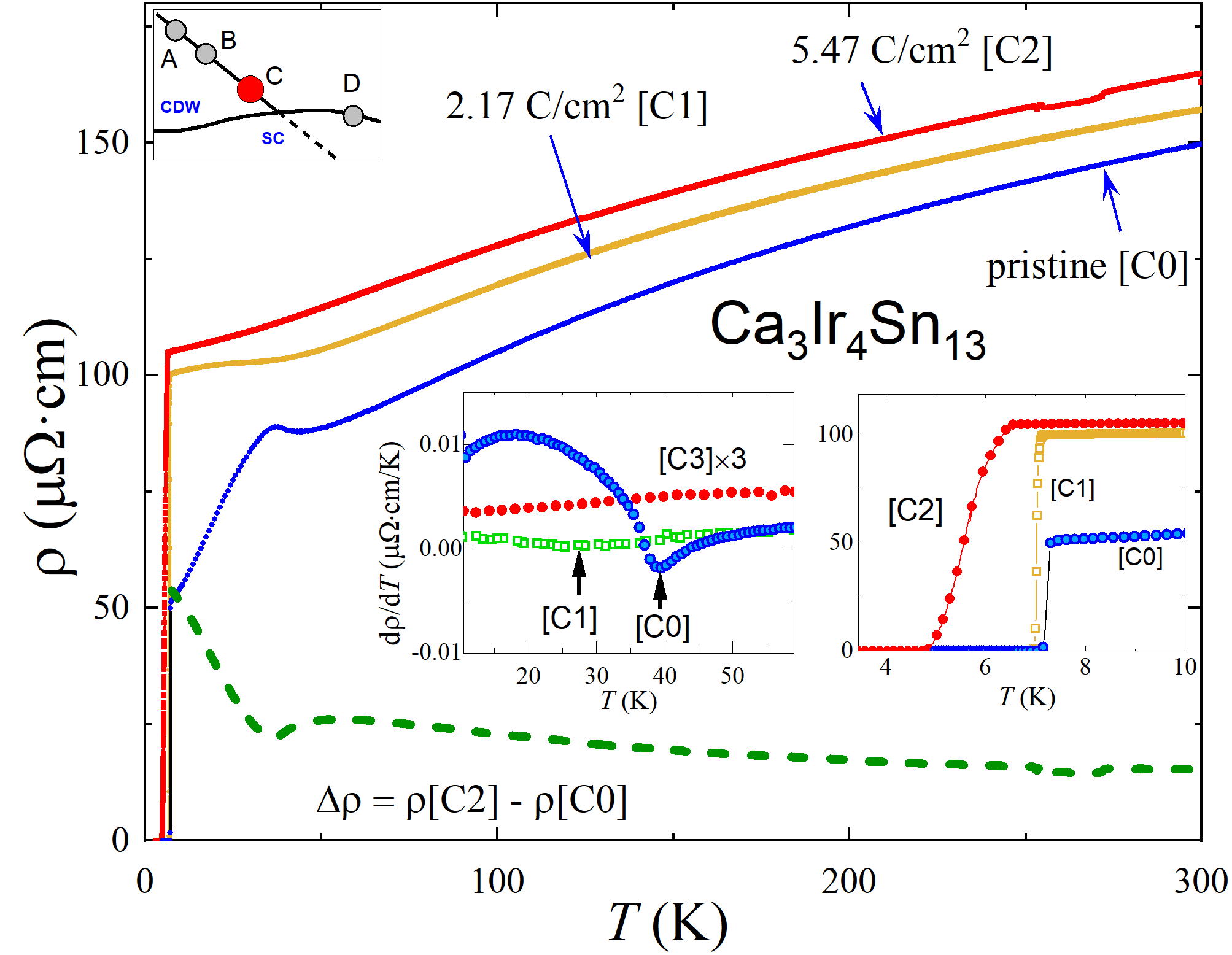} 
\caption{Temperature-dependent resistivity of $\text{Ca}_{3}\text{Ir}_{4}\text{Sn}_{13}$
before irradiation (blue line), after receiving 2.17 C/cm$^{2}$ of
irradiation (yellow line), and then an additional 3.3 C/cm$^{2}$
for a total dose of 5.47 C/cm$^{2}$ of electron irradiation. The
green dashed line shows the difference between the pristine and 5.47
C/cm$^{2}$ curves, showing deviation from Matthiessen's rule below
the transition temperature $T_{\text{CDW}}$. The small cartoon in
the top left corner indicates sample position on generic phase diagram.
The left inset shows the derivative of the resistivity showing suppression
and blurring of the CDW phase transition with irradiation. The right
inset shows the shift in the superconducting transition temperature. }
\label{fig:Ca3Ir4Sn13} 
\end{figure}

$\text{Ca}_{3}\text{Ir}_{4}\text{Sn}_{13}$ is the compound with the
lowest $T_{\text{CDW}}$. As shown in Fig.~\ref{fig:Ca3Ir4Sn13},
a clear feature in the temperature-dependent resistivity is observed
at $\sim40\:K$ in the pristine sample (arrow in the derivative plot,
left inset). It is also the closest CDW composition to the structural
quantum critical point. The suppression of CDW with irradiation is
clear for 2.17 C/cm$^{2}$ irradiation. The transition feature in
the derivative plot cannot be resolved at 5.47 C/cm$^{2}$, suggesting
that the CDW order has been completely suppressed. Similar to $\text{Sr}_{3}\text{Ir}_{4}\text{Sn}_{13}$,
but unlike $\text{Sr}_{3}\text{Rh}_{4}\text{Sn}_{13}$ increasing
the disorder in this compound only decreases the superconducting transition
temperature Furthermore, unlike the $\text{Sr}_{3}\text{(Ir,Rh)}_{4}\text{Sn}_{13}$
compounds, Matthiessen's rule is weakly violated in this material
for all temperatures, below and above $T_{\text{CDW}}$. Matthiessen's
rule holds in good metals, where the introduction of disorder affects
only the residual resistivity (scattering off the defects and impurities)
and appears as a constant offset. Since the change of resistivity
under increasing disorder in $\text{Ca}_{3}\text{Ir}_{4}\text{Sn}_{13}$
is more complex than just a constant offset, it may suggest the presence
of a short-range order consistent with the above discussion and similarity
with $\text{2H-NbSe}_{2}$ \cite{NbSe2disorder}.

\subsection{$\text{Ca}_{3}\text{Rh}_{4}\text{Sn}_{13}$}

\begin{figure}[tb]
\centering
\includegraphics[width=0.9\linewidth]{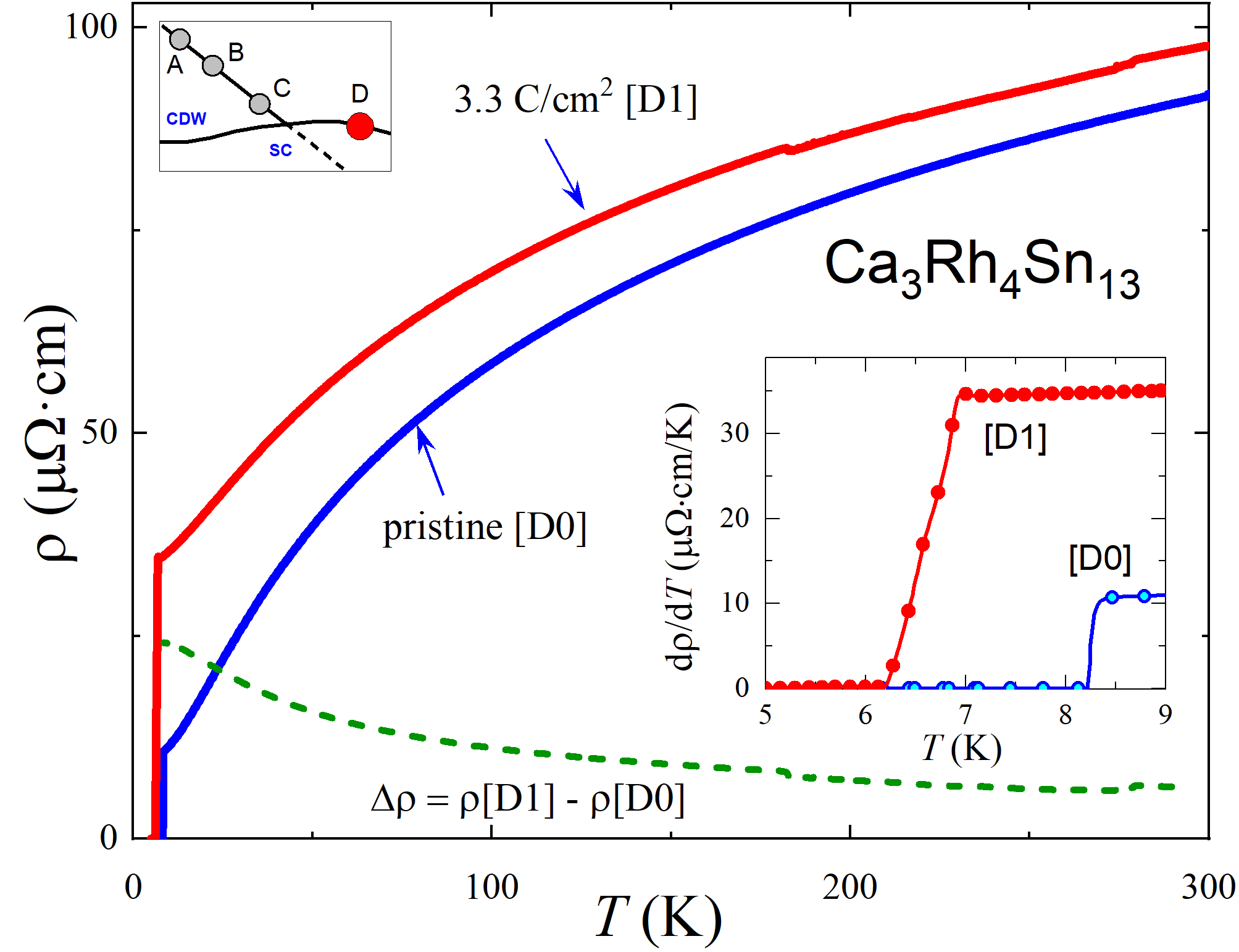} 
\caption{Temperature-dependent resistivity of $\text{Ca}_{3}\text{Rh}_{4}\text{Sn}_{13}$
in the pristine state before irradiation (blue curve) and after 2.1
C/cm$^{2}$ irradiation (red curve). The green dashed line shows the
difference between the curves, finding strong Matthiessen's rule violation
at all temperatures above the superconducting transition. The small
cartoon in the top left corner indicates the position of the compound
on the generic phase diagram, - the right-most, with no CDW transition.
The inset shows resistivity in the vicinity of the superconducting
transition, revealing substantial $T_{c}$ suppression by electron
irradiation. }
\label{fig:Ca3Rh4Sn13} 
\end{figure}

$\text{Ca}_{3}\text{Rh}_{4}\text{Sn}_{13}$ is our only compound which
does not have a long-range CDW ordering, and is positioned to the
right of the quantum critical point in the generic phase diagram in
Fig.~\ref{fig:PhaseDiagram}(a). Still, the evolution of the temperature-dependent
resistivity with disorder, Fig.~\ref{fig:Ca3Rh4Sn13}, reveals that
Matthiessen's rule is conspicuously \textit{not} obeyed in the ``normal''
state. Similar to $\text{Sr}_{3}\text{Ir}_{4}\text{Sn}_{13}$, this
suggests that there is some other type of (short-range) electronic
order which is affected by the introduction of point-like disorder.
One potential candidate is the residual short-range CDW order which persisted
across the QCP, as observed in 2H-NbSe2 \cite{NbSe2disorder}.  We note that second-order structural phase transition in a 3-4-13 family, specifically $\text{La,Ce}_{3}\text{Rh}_{4}\text{Sn}_{13}$, has been discussed in the context of unconventional chiral CDW based on structural X-ray studies \cite{Suyama2018}.

The superconducting transition temperature in $\text{Ca}_{3}\text{Rh}_{4}\text{Sn}_{13}$
is significantly affected by electron irradiation. $T_{c}$ is suppressed
from $T_{c,0}=8.2$~K by more than 2~K after 2.1 C/cm$^{2}$ of irradiation using the zero-resistance
offset criterion, and the transition broadens. Potential scattering
is not expected to suppress $T_{c}$ in conventional isotropic single-band
$s-$wave superconductors, so we must consider the possibility of
nodal superconductivity, or at least a strong variation of the superconducting
gap magnitude on the Fermi surfaces. This will be addressed in detail
in the next section.

\section{Discussion\label{Sec:Discussion}}

\begin{figure}[tb]
\centering
\includegraphics[width=0.9\linewidth]{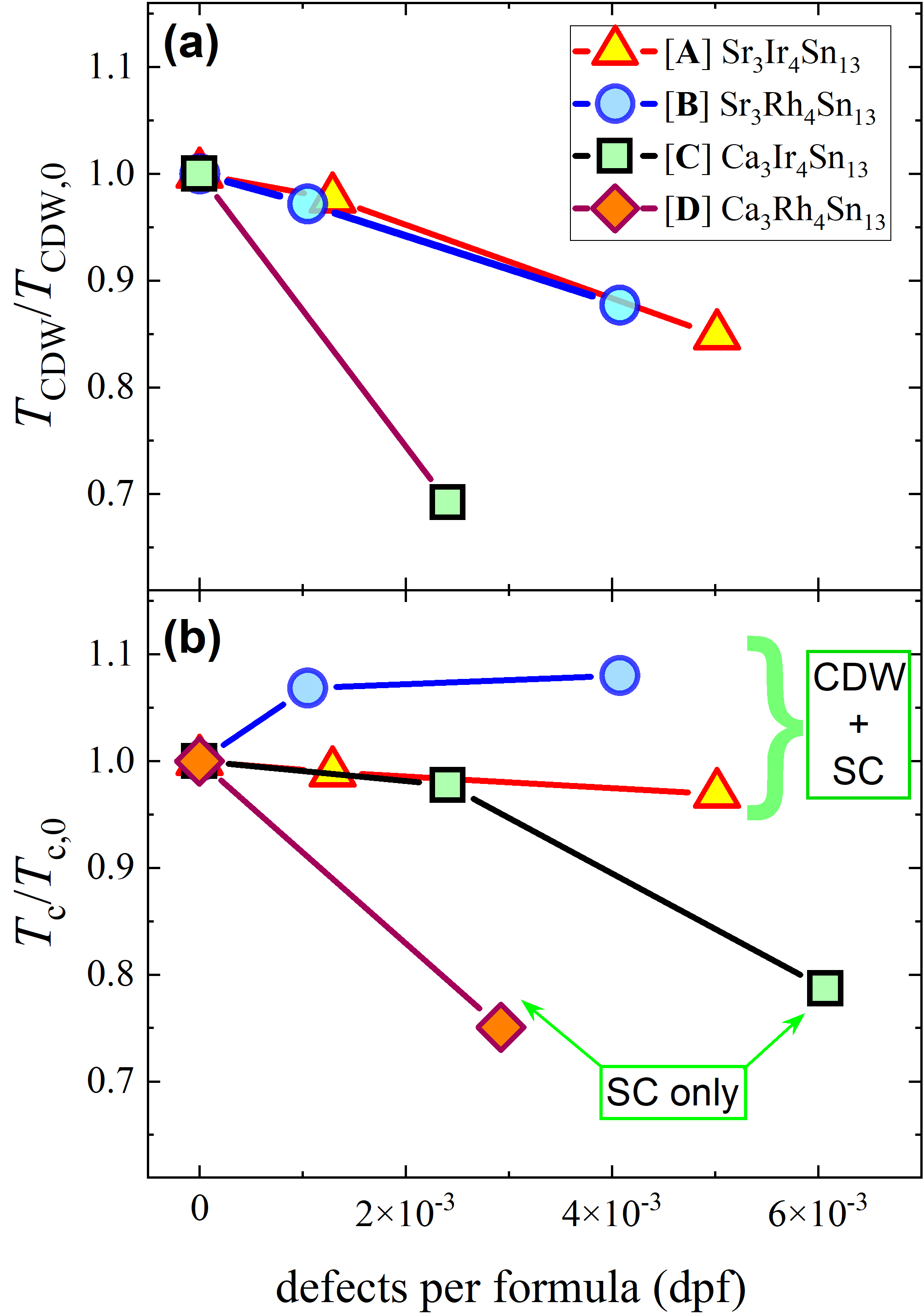} 
\caption{(a) The evolution of the CDW transition temperature normalized by
the value before irradiation, $T_{\text{CDW}}/T_{CDW,0}$, and (b)
similar plot of the superconducting transition temperature normalized
by the value in pristine samples, $T_{c}/T_{c,0}$ with the induced
disorder in units of defects per formula unit (dpf), see text for
details. The $X$ and $Y$ scales on both graphs are the same for easy comparison. Note a significant increase of $T_c$ suppression in samples where CDW does not coexist with superconductivity.}
\label{fig:SummaryFigure} 
\end{figure}

The increase of a sample's residual resistivity as a function of irradiation
dose is an intrinsic measure of the disorder introduced by irradiation.
However, because the resistivity in the CDW state depends on the size
of the gapped part of the Fermi surface, which is compound-dependent,
a direct comparison across chemical compositions is not very informative.
A better proxy for the quantification of the effect of disorder across
compounds is the resistivity in the metallic state above $T_{\text{CDW}}$.
As can be seen from Fig.~\ref{fig:PhaseDiagram}(b), the $\rho(T)$
curves for all compounds are nearly parallel approaching room temperature,
and the overall resistivity variation does not exceed $30\:\%$ or
so. The values of $\rho\left(300\:K\right)$ are listed in Table \ref{tab:Normal-state-resistivity}.
This validates the assumption of practically negligible differences
in the carrier density between the different compounds in the normal
state at elevated temperatures. By measuring the change in the resistivity
in the normal state, we can thus determine the change in the disorder
scattering. Combined with the numerical estimates of the defect density
as described in Section \ref{ExpDetails}, these measures allow for
a direct comparison between different samples.

\subsection{Interplay of charge-density wave and superconductivity}

In the following, we summarize critical temperatures extracted from
figures 3-6 as a function of irradiation. Ideally, such summary plots would show the error bars in both $X$ and $Y$ directions. However, it cannot be done in our case because we did not measure many samples that would allow statistical analysis. On the other hand, each critical temperature is determined with such precision that the corresponding uncertainty error bar is smaller than the symbol size. This is also true for estimating the $X$ axis values that involve measured total irradiation dose and residual resistivity.

Figure ~\ref{fig:SummaryFigure}(a) shows the evolution of the CDW
ordering temperature, $T_{\text{CDW}}$, with the defects per formula
unit (dpf). The observed dependence is striking. While the CDW is
suppressed at the same rate in $\text{Sr}_{3}\text{(Ir,Rh)}_{4}\text{Sn}_{13}$
compounds, the closer to QCP $\text{Ca}_{3}\text{Ir}_{4}\text{Sn}_{13}$
shows a much larger suppression rate. Intriguingly, this is the composition
where the Matthiessen's rule is violated above the CDW transition, as it
is expected that quantum fluctuations affect the properties near QCP.
A similar graph of the normalized $T_{c}/T_{c,0}$ in Fig.~\ref{fig:SummaryFigure}(b)
shows complex behavior. The rate of suppression is similar in samples
{[}A{]} and {[}C{]}, while  $T_{c}$ increases in sample {[}B{]}. Such increase is expected when superconducting pairing and charge-density
wave inter-band interaction energies are comparable and the enhancement of superconductivity due to CDW suppression over-weights the natural suppression of $T_c$ by disorder. For $\text{Ca}_{3}\text{Rh}_{4}\text{Sn}_{13}$
which is away from CDW phase, the suppression of $T_{c}$ is dramatic,
despite the fact that its $T_{c,0}$ is similar to {[}B{]} and {[}C{]}. Interestingly, and consistent with this picture as soon as CDW is completely suppressed in sample {[}C{]}, the $T_c$ suppression becomes much faster and similar to sample {[}D{]}. Our measurements establish that in 3-4-13 stannides, there is a direct competition of CDW and superconductivity, in addition to quantum fluctuations around QCP that affect even normal-state properties. Of course, despite similarities, we are dealing with four distinctly
different compounds and some unique structural and/or electronic features
may certainly contribute to the results.

\begin{figure}[tb]
\centering
\includegraphics[width=0.9\linewidth]{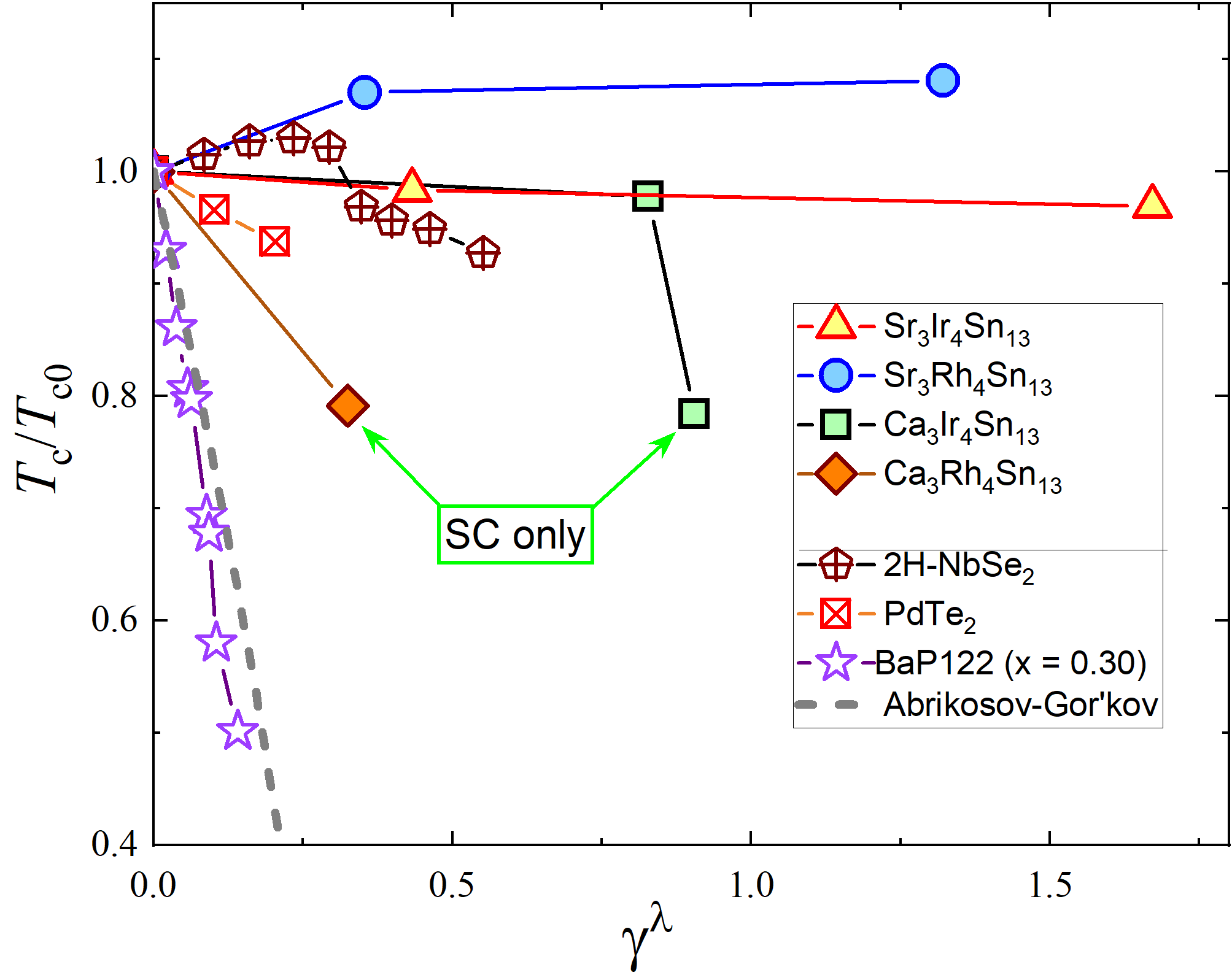} 
\caption{ Normalized suppression of the superconducting transition temperature, $T_{c}/T_{c0}$, as function of the dimensionless scattering rate $\gamma^{\lambda}$ evaluated from resistivity and London penetration depth, $\lambda\left(0\right)$
using Eq.\eqref{eq:gamma-lambda}. To set the scale, grey dashed line shows the predictions of Abrikosov-Gor'kov theory for a line-nodal $d$-wave superconductor with non-magnetic impurities \cite{AG,Openov1997,Openov2004}. Purple stars are experimental data for the nodal iron-based superconductor BaFe$_{2}$(As$_{1-x}$P$_{x}$)$_{2}$ \cite{Hashimoto2012},
red-yellow squares are for the superconducting Dirac semimetal PdTe$_{2}$,
maroon cross-pentagons are for CDW superconductor $\text{2H-NbSe}_{2}$. The
data for 3-4-13 compounds are shown by red rhombi for $\text{Ca}_{3}\text{Rh}_{4}\text{Sn}_{13}$,
blue squares for $\text{Ca}_{3}\text{Ir}_{4}\text{Sn}_{13}$, yellow-blue
circles for $\text{Sr}_{3}\text{Rh}_{4}\text{Sn}_{13}$ and yellow
triangles for $\text{Sr}_{3}\text{Ir}_{4}\text{Sn}_{13}$.}
\label{fig:scattering} 
\end{figure}

\subsection{Matthiessen's rule}

The temperature-dependent resistivity in the normal state of the 3-4-13
system is anomalous and reveals notable Matthiessen's rule violations
in the vicinity of the QCP, but not too far away. 
Comparison of the four compounds finds some similarity at high temperatures. At temperatures above massive downturn in the resistivity on cooling, coinciding with $T_{CDW}$ in $\text{Sr}_{3}\text{Ir}_{4}\text{Sn}_{13}$ and $\text{Sr}_{3}\text{Rh}_{4}\text{Sn}_{13}$, the $\rho(T)$ curves extrapolate to quite high values in $T=0$ limit. This feature is known to be caused by spin-disorder scattering in magnetic materials \cite{spindisorder}. It is also observed above $T_{CDW}$ in tantalum dichalcogenides, TaS$_2$ and TaSe$_2$ \cite{Naito1982,Naito2nd} and was suggested to be scattering on charge fluctuations above the transition. In $\text{Ca}_{3}\text{Ir}_{4}\text{Sn}_{13}$ and particularly strongly in  $\text{Ca}_{3}\text{Rh}_{4}\text{Sn}_{13}$, the low-temperature downturn in $\rho (T)$ does not coincide with long range CDW ordering. This type of response may be suggestive of the scenario realised in NbSe$_2$ \cite{NbSe2disorder}. Here long range charge density wave ordering is suppressed with irradiation, however short range ordering remains unaffected.

It is interesting
to compare this behavior with another fully gapped system where doping-dependent
spin density (SDW) wave coexists with superconductivity, for example
electron-doped Ba(Fe$_{1-x}$Co$_{x}$)$_{2}$As$_{2}$ (BaCo122)
\cite{BaKMathesson,Joshi2020} and iso-electron-substituted iron-based
superconductor BaFe$_{2}$(As$_{1-x}$P$_{x}$)$_{2}$ \cite{Shibauchidisorder},
both showing proven SDW/QCP under the dome of superconductivity \cite{Hashimoto2012,Wang2018,Joshi2020}.
In these compounds, Matthiessen's rule is obeyed near the QCP,
as well as it is in the cuprates if the sample is not in the regime
of weak localization \cite{tccuprates}. On the other hand, the observed
behavior of 3-4-13 bears some similarity to the hole doped $\text{Ba}_{1-x}\text{K}_{x}\text{Fe}_{2}\text{As}_{2}$
(BaK122) in which Matthiessen's rule is also strongly violated \cite{BaKMathesson}. 

\subsection{Dimensionless scattering rate}

To put our data in a broader perspective, we compare the $T_{c}$
suppression rate in the 3-4-13 compounds with other known cases. For this, we will use a dimensionless scattering rate defined as 
defined as~\cite{PRX, TimmonsPdTe2}
\begin{equation}
\gamma^{\lambda}=\frac{\hbar \Delta \tau^{-1}}{2\pi k_{B}T_{c,0}}=\frac{\hbar}{2\pi k_{B}\mu_{0}}\frac{\Delta\rho_{0}}{\lambda_{\text{clean}}^{2}(0)T_{c,0}}\,.
\label{eq:gamma-lambda}
\end{equation}
\noindent Here, $\Delta\rho_{0}$ is the change of the residual resistivity after irradiation compared to the pristine state value, and $\lambda_{\text{clean}}(0)$ is the zero temperature penetration depth in the pristine sample. Note that we obtain $\Delta \rho_0$ by extrapolation to $T=0$. Inserting the dimensional constants and using units of $\mu \Omega \text{cm}$ for $\Delta \rho_0$, $10^{-7}~\text{m}$ for the penetration depth $\lambda_{\text{clean}}(0)$, and K for $T_{c,0}$, Eq.~\ref{eq:gamma-lambda} takes the form $\gamma^\lambda = 0.97 \Delta \rho_0/\left(\lambda_\text{clean}^2(0) T_{c,0}\right)$.

To arrive at Eq.~\eqref{eq:gamma-lambda} we used the simple Drude model for resistivity, $\rho =m^*/(n e^2 \tau)$, and the London model for the penetration depth, $\lambda_{\text{clean}}^{2}(0) =m^*/\left( \mu_0 n e^2 \right)$~\cite{Tinkham_book} (see also Appendix D of Ref.~\cite{TimmonsPdTe2}). Note that we have used that the superfluid density equals the total carrier density at zero temperature. This allows expressing the (change of the) normal-metal scattering time via measurable parameters, $\Delta \tau^{-1}=\Delta\rho_{0}/\mu_{0}\lambda_{\text{clean}}^{2}(0)$. We note that $\lambda_{\text{clean}}(0)$ and the normal-state scattering time, $\tau$, do not depend on parameters of superconductivity and Eq.~\eqref{eq:gamma-lambda} can thus be used for different gap symmetries. 

Now we can compare the results of 3-4-14 stannides with various theoretical
expectations as well as other superconductors in which the effect
of disorder was studied. Figure \ref{fig:scattering} shows normalized
$T_{c}$ suppression for our four systems as a function of $\gamma^{\lambda}$.
The data are compared with nodal $s^{\pm}$ BaFe$_{2}$(As$_{1-x}$P$_{x}$)$_{2}$,
\cite{BaP}, Dirac semi-metal PdTe$_{2}$~\cite{TimmonsPdTe2}, and
CDW superconductor $\text{2H-NbSe}_{2}$~\cite{NbSe2disorder}. The expectation
from the Abrikosov-Gor'kov theory for a single-band $d$-wave superconductor
with non-magnetic scattering \cite{AG,Openov1997,Openov2004}, shown
by the dashed line, provides the scale for the largest suppression rate possible. While in three CDW/SC 3-4-13 compounds, it can be argued that anything is possible due to cooperation and/or competition between
these two quantum orders, the significant $T_{c}$ suppression rate
in $\text{Ca}_{3}\text{Rh}_{4}\text{Sn}_{13}$ is shown to be intermediate
between nodal and nodeless superconductors. In fact, it is comparable
to the $T_{c}$ suppression rate in the nodeless sign-changing $s^{+-}$
state of the optimally doped $\text{Ba}(\text{Fe}_{1-x}\text{Ru}_{x})_{2}\text{As}_{2}$~\cite{PRX},
and is significantly higher than that of a two-gap $s^{++}$ $\text{2H-NbSe}_{2}$
after the suppression of CDW order. This relatively high $T_{c}$
suppression rate naturally raises questions about the superconducting
gap structure of $\text{Ca}_{3}\text{Rh}_{4}\text{Sn}_{13}$ and to
get an insight into the momentum dependence of the order parameter,
we measured the London penetration depth in $\text{Ca}_{3}\text{Rh}_{4}\text{Sn}_{13}$.

\subsection{London penetration depth of $\text{Ca}_{3}\text{Rh}_{4}\text{Sn}_{13}$\label{LPD}}

\begin{figure}[tb]
\centering
\includegraphics[width=1\linewidth]{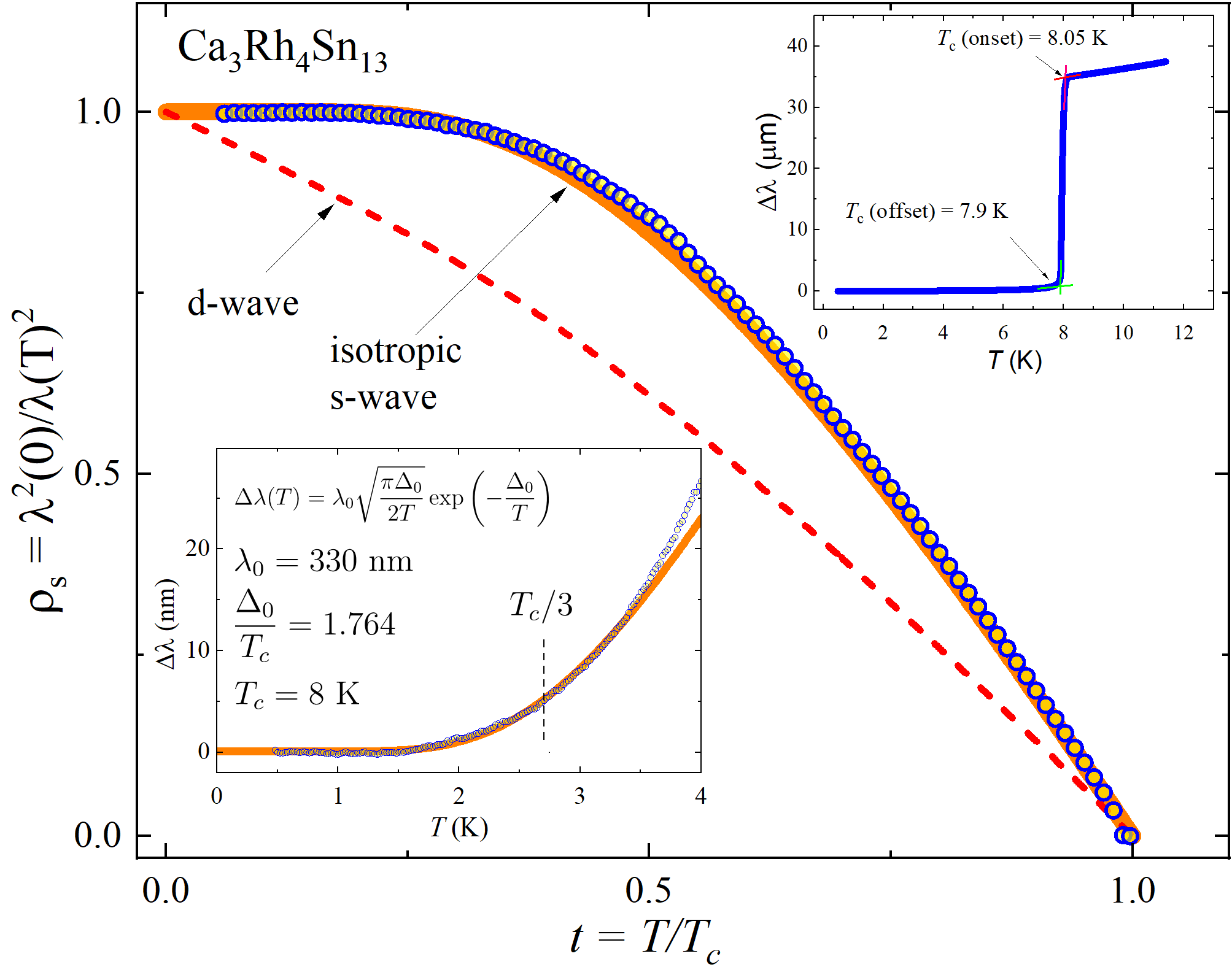} 
\caption{Main panel: superfluid density in $\text{Ca}_{3}\text{Rh}_{4}\text{Sn}_{13}$
calculated from the London penetration depth, $\lambda\left(T\right)$,
measured using the tunnel-diode resonator technique. The thick orange
line shows standard isotropic $s-$wave BCS behavior, while the dashed
line shows the expectation for a nodal $d-$wave order parameter.
The lower inset shows a BCS fit of $\Delta\lambda\left(T\right)$
to the single-gap expression shown. The only fitting parameter was
$\lambda(0)=330$ nm, while the $T_{c}$ and the weak-coupling gap
ratio, $\Delta_{0}/T_{c}=1.764$ was fixed. The obtained $\lambda(0)=330\:\text{nm}$
was used to construct the $\rho_{s}(T)=\left(1+\Delta\lambda(T)/\lambda(0)\right)^{-2}$
shown in the main panel. The upper inset shows the sharp transition
of our high-quality sample and $\lambda(T_{c})$ consistent with the
expected skin depth of the normal state. }
\label{fig:lambda} 
\end{figure}

To examine the anisotropy of the energy gap, we used a sensitive tunnel-diode
resonator technique, described in the experimental methods, \secref{ExpDetails}, to measure the low-temperature variation of the
London penetration depth in $\text{Ca}_{3}\text{Rh}_{4}\text{Sn}_{13}$.
Figure \ref{fig:lambda} shows the variation of the superfluid density,
$\rho_{s}=\lambda^{2}(0)/\lambda^{2}(T)$, calculated from the measured
variation of the London penetration depth, $\Delta\lambda(T)=\lambda(T)-\lambda(0)$,
over the whole temperature range. This is important to detect possible
signatures of two-gap superconductivity. The top right inset in Fig.~\ref{fig:lambda}
shows a full-range variation of $\Delta\lambda(T)$ and the lower
left inset zooms on the characteristic low-temperature range, approximately
$T<T_{c}/3$, where the order parameter amplitude is practically constant
and any changes in $\lambda(T)$ come from the quasiparticles generated
due to angular variation of the gap function. The red line in the
bottom-left inset shows an excellent fit to the isotropic single-gap
$s$-wave function with $\lambda(0)=330\,\textrm{nm}$ and $\Delta_{0}/k_{B}T_{c}=1.764$. There are no reported measurements of $\lambda(0)$ in $\text{Ca}_{3}\text{Rh}_{4}\text{Sn}_{13}$, however $\mu\text{SR}$ measurements report $\lambda(0)=291\,\textrm{nm}$ in $\text{Sr}_{3}\text{Ir}_{4}\text{Sn}_{13}$ \cite{multibandmusR} and $\lambda(0)=351\,\textrm{nm}$ in $\text{Ca}_{3}\text{Ir}_{4}\text{Sn}_{13}$ \cite{Biswas2014}; so our measurement is perfectly in range considering that $\lambda(0)$ is a normal-state property that depends only on the parameters of electronic band-structure.
The superfluid density calculated from the obtained $\lambda(T)=\lambda(0)+\Delta\lambda(T)$  (main
panel, symbols) is in very good agreement with the parameter-free
prediction for an isotropic full-gap $s$-wave superconducting state
(main panel, thick orange line). For comparison, the expectation for
a $d-$wave superconductor is shown by the dashed line. This nearly
perfect and robust agreement with the simplest isotropic BCS is at
odds with the significant rate of the disorder-induced reduction of
the $T_{c}$. As we show below, these conclusions are impossible to
reconcile without invoking unconventional pairing in $\text{Ca}_{3}\text{Rh}_{4}\text{Sn}_{13}$.

Measurements of $\lambda(T)$ allow us to address the question of whether the defects induced by electron irradiation become magnetic. In principle, non-magnetic ions may become magnetic when their ionization changes. Such magnetic defects can cause pair-breaking due to spin-flip scattering, resulting in a $T_c$ suppression even in isotropic $s-$wave fully-gapped superconductors  \cite{AG,Openov1997,Openov2004}.
Our precision measurements of the London penetration depth in this system exclude this scenario.  Due to the sensitivity of these measurements, even a minute paramagnetic signal coming from magnetic defects would be detected. In particular, in the presence of magnetic impurities, the London penetration depth estimated from the magnetic susceptibility measurements (such as tunnel-diode resonator) is renormalized as $\lambda_{m}\left(T\right)=\sqrt{\mu\left(T\right)}\lambda\left(T\right)$,
where $\lambda\left(T\right)$ is the London penetration depth of a superconducting
sample without magnetic impurities and $\mu\left(T\right)$ is the normal-state
magnetic permeability due to dilute non-interacting magnetic moments
(ions, impurities etc). We refer to Ref.~\cite{Prozorov2000}, which shows 
what the measured penetration depth looks like when this effect is relevant.  
Here we do not see any trace of the paramagnetic upturn expected if we had magnetic impurities.
From the concentration of defects induced by irradiation, which is up to
$5\times10^{-3}$ dpf, the volume of the conventional unit cell, $\left(9.7\:\text{Å}\right)^{3}$,
one obtains with $Z=2$ formulas for the the concentration of defects in conventional
units $n_{d}\approx1\times10^{25}\:\text{m}^{-3}=18\:\text{mol}/\text{m}^{3}$.
Now we can evaluate the Curie constant.  Assuming the simplest case, that
each scattering center is a two-level system with the magnetic moment
of one Bohr magneton, $\mu=\mu_{B}=9.27\times10{{}^-}{{}^2}{{}^4}\,\mathrm{J/T}$.
With the estimated $n_{d}$ we obtain, $C=\mu_{0}n_{d}\mu_{B}^{2}/k_{B}\approx7.8\times10{{}^-}{{}^5}\:\mathrm{K}.$
This is a very small number even for such a large moment. It gives
a correction to our penetration depth, $\lambda\left(0\right)=330\:\text{nm}$, of
approximately $\Delta\lambda\left(0.4\:\mathrm{K}\right)\approx0.32\:\text{\AA}$
at the minimum temperature of 0.4 K. This is a negligible correction.
Of course, when $T\rightarrow0$, it will grow large, but in this paper we are mostly examining what happens at $T_{c}$, and such a dilute system will not
be able to shift $T_{c}$ in any appreciable way. If for some
reason the magnetic moment is larger or more defects are generated, the
measurements of London penetration depth are capable of resolving
sub-nm variation and would pick up such a signal. We can therefore say
with confidence that magnetism of the defects induced by electron irradiation
does not play a role in the obtained results. 

Finally, together with a very sharp resistive and magnetic transitions in pristine sample, the behavior of  $\lambda(T)$  also rules out possible chemical and structural inhomogeneities that were shown to lead to a significant spread of the observed $T_c$ in polycrystalline  $\text{Ca}_{3}\text{Rh}_{4}\text{Sn}_{13}$ \cite{Slebarski2016}.

\subsection{Candidate pairing states for $\text{Ca}_{3}\text{Rh}_{4}\text{Sn}_{13}$}

Since $\text{Ca}_{3}\text{Rh}_{4}\text{Sn}_{13}$ does not exhibit
a transition into a CDW phase, the normal state symmetries out of
which superconductivity emerges are expected to be those of the room-temperature
symmetry group of the 3-4-13 series---the space group $Pm\bar{3}n$
(no.~223) with associated point group $O_{h}$; this is confirmed
by XRD measurements \cite{Slebarski2018}. Both in the literature
and in our measurements, there are no indications of multiple consecutive
superconducting transitions. Therefore, we can use the irreducible representations (IRs) of the normal-state symmetry group \cite{SigristUeda} to classify the superconducting
order parameters. In the absence
of magnetic fields, it is further natural to assume that the pairing
order parameter transforms trivially under lattice translations and
we can focus on the IRs of the point group $O_{h}$. Note that the
involved atoms are moderately heavy and we thus expect spin-orbit
coupling to be sufficiently strong such that the symmetries of O$_{h}$
should be thought of as acting on the spatial coordinates (three-dimensional
momentum $\vec{k}$) and spin simultaneously.

\global\long\def\arraystretch{1.2}%
 
\begin{table*}[tb]
\begin{centering}
\caption{Possible fully gapped pairing states for $\text{Ca}_{3}\text{Rh}_{4}\text{Sn}_{13}$ as constrained by
the point group O$_{h}$. The first four states above the horizontal
line can be fully gapped right below the superconducting critical
temperature $T_{c}$. The two states below the horizontal line exhibit
line nodes right below $T_{c}$ but can, in principle, be fully gapped
at sufficiently low $T$. The column $d_{n}$ denotes the dimensionality
of the IR and the column TRS states whether the pairing state has
time-reversal symmetry. We use the short-hand notation $X=X_{\vec{k}}$,
$Y=Y_{\vec{k}}$, $Z=Z_{\vec{k}}$ to denote real-valued Brillouin-zone-periodic
functions that transform as $x$, $y$, and $z$ under O$_{h}$. We
also indicate the ratio of the maximal to minimal value of the superconducting
gap, $\Delta_{\text{max}}/\Delta_{\text{min}}$, for an isotropic
Fermi surface around $\vec{k}=0$ and with $(X,Y,Z)=(k_{x},k_{y},k_{z})$.
As discussed in the text, only the states of ``type'' (ia) are natural candidates consistent with the temperature dependence of the superfluid density in \figref{fig:lambda}.}
\label{PossiblePairingStates} 
\end{centering}
\begin{ruledtabular}
\begin{tabular}{ccccccc}
IR  & pairing  & $d_{n}$  & TRS  & order parameter $\Delta_{\vec{k}}i\sigma_{y}$  & $\Delta_{\text{max}}/\Delta_{\text{min}}$  & type \tabularnewline
\hline 
$A_{1g}$  & $s$ wave  & $1$  & \cmark  & $a+b(X^{2}+Y^{2}+Z^{2})$  & $1$  & (ia) \tabularnewline
$A_{1u}$  & $p$ wave  & $1$  & \cmark  & $X\sigma_{x}+Y\sigma_{y}+Z\sigma_{z}$  & $1$  & (ia) \tabularnewline
$E_{u}$  & $e_{u(0,1)}$ wave  & $2$  & \cmark  & $2Z\sigma_{z}-X\sigma_{x}-Y\sigma_{y}$  & $2$  & (ib) \tabularnewline
$T_{2u}$  & $t_{2u(1,1,1)}$ wave  & $3$  & \cmark  & $(Y+Z)\sigma_{x}+(X+Z)\sigma_{y}+(X+Y)\sigma_{z}$  & $2$  & (ib) \tabularnewline
\hline 
$E_{g}$  & $e_{g(0,1)}$ wave  & $2$  & \cmark  & $2Z^{2}-X^{2}-Y^{2}$  & $\infty$ (line nodes)  & (ii) \tabularnewline
$T_{2g}$  & $t_{2g(1,1,1)}$ wave  & $3$  & \cmark  & $YZ+ZX+XY$  & $\infty$ (line nodes)  & (ii) \tabularnewline
\end{tabular}
\end{ruledtabular}

\end{table*}

Since O$_{h}$ contains inversion, $i$, all bands are doubly-degenerate
despite the presence of spin-orbit coupling. We label the degeneracy
with a pseudospin quantum number. Another consequence of $i\in O_{h}$
is that all IRs decay into even, $g$, and odd, $u$, representations
under $i$, associated with pseudospin singlet and triplet. For each
$\mu=g,u$, O$_{h}$ has two 1D IRs, $A_{1\mu}$ and $A_{2\mu}$,
one 2D IR, $E_{\mu}$, and two 3D IRs, $T_{1\mu}$ and $T_{2\mu}$,
leading to a total number of $10$ IRs. This gives rise to a large
number (26) of possible pairing states~\cite{SigristUeda}. However,
most of these states necessarily have nodes which is not consistent
with the observed temperature dependence of the penetration depth
in Fig.~\ref{fig:lambda}. As summarized in \tableref{PossiblePairingStates},
only six states are left that \textit{can} be fully gapped. When specifying
the superconducting order parameter, $\Delta_{\vec{k}}$, in \tableref{PossiblePairingStates},
we focus on generic momenta on the Fermi surfaces without additional
degeneracies between different bands. Therefore, $\Delta_{\vec{k}}$
can be taken to be a $2\times2$ matrix in pseudospin-space, which
we have expanded in terms of Pauli matrices $\sigma_{j}$ in \tableref{PossiblePairingStates}.

These six candidate states can be further divided into two categories:
(i) four states that will be fully gapped right below $T_{c}$ since
their primary order parameters are associated with a full gap: these
are the regular BCS $s$-wave, spin singlet state, transforming under
$A_{1g}$, a helical triplet ($A_{1u}$), and two triplets transforming
under $E_{u}$ and $T_{2u}$, respectively; (ii) two states where
the primary order parameter has line nodes but which, once non-zero,
can induce secondary superconducting orders that have a full gap:
these are two singlets, one transforming under $E_{g}$ and one under
$T_{2g}$.

The states (ii) are not consistent with experiment for the following
reasons: they will have line nodes in a finite range below $T_{c}$,
which together with the temperature-dependent admixture of a secondary
order parameter, is generically expected to lead to a more unconventional
temperature dependence of the penetration depth than what is seen
in Fig.~\ref{fig:lambda}. Further, the admixture of the secondary
component has to be extremely large to not only remove the nodes but
also lead to an approximately isotropic gap function (see also \appref{Admixture}).

Among the remaining four states of type (i) in \tableref{PossiblePairingStates},
we can further distinguish between (ia) states that can have a fully
isotropic gap function and (ib) states which are, by symmetry, forced
to have a momentum-dependent order parameter that generically leads
to a significantly momentum-dependent gap. The ratio of the maximum
to minimum gap size, $\Delta_{\text{max}}/\Delta_{\text{min}}$, on
the Fermi surface is expected to be of the order of $2$ for the (ib)
states. Based on the penetration-depth data, the (ia) states thus
seem more natural candidates. We therefore focus for the following
analysis of the irradiation data on the $A_{1g}$ singlet and $A_{1u}$
triplet states.

\subsection{Constraints on pairing from sensitivity to disorder scattering}
\label{ConstraintOnPairingFromDisorder} 

To quantitatively analyze the measured impact of impurities on $T_{c}$ in $\text{Ca}_{3}\text{Rh}_{4}\text{Sn}_{13}$, we
use the general expression derived in \refcite{TimmonsPdTe2} for
the sensitivity parameter $\zeta$ that describes the disorder-induced
reduction of the superconducting critical temperature according to

\begin{equation}
\frac{T_{c,0}-T_{c}(\tau^{-1})}{T_{c,0}}\sim\frac{\pi}{4T_{c,0}}\tau^{-1}\,\zeta.\label{eq:Tcsuppression}
\end{equation}
This expression holds in the limit of weak scattering rates, $\tau^{-1}\rightarrow0$,
where $\zeta$ corresponds to the linear slope of the $T_{c}$ reduction
as a function of $\tau^{-1}$. With the normalization in \equref{eq:Tcsuppression},
we have $\zeta=1$ for magnetic impurities in a single-band, isotropic,
spin-singlet superconductor and $\zeta=1/2$ for a single-band $d$-wave
superconductor in the presence of non-magnetic impurities (see grey
dashed line in Fig.~\ref{fig:scattering}). Comparison of the slopes
in Fig.~\ref{fig:scattering} allows to extract $\zeta\approx1/9$
for our $\text{Ca}_{3}\text{Rh}_{4}\text{Sn}_{13}$ sample. \refcite{TimmonsPdTe2} related $\zeta$
for a general superconductor and a general form of disorder to a (properly
normalized) Frobenius norm of the commutator appearing in the generalized
Anderson theorem of Refs.~\cite{Scheurer2016,Hoyer2015}. The full
expression for general disorder potentials and pairing states is defined
in \appref{ExpressionFoZeta} {[}see~\equref{ZetaExpression}{]}.
In the following, we apply it to the relevant pairing states in $\text{Ca}_{3}\text{Rh}_{4}\text{Sn}_{13}$
that were identified above. Since electron irradiation creates point-like,
non-magnetic defects, we focus on this type of disorder.

We begin with the $A_{1g}$ singlet and assume a general momentum-dependent
order parameter, $\Delta_{\vec{k}}=\Psi_{\vec{k}}i\sigma_{y}$ where
$\Psi_{\vec{k}}$ is invariant under all symmetries of $O_{h}$. 
Considering point-like, non-magnetic disorder without any momentum
dependence in the pseudospin basis, \equref{ZetaExpression} readily
yields 
\begin{equation}
\zeta=\frac{\braket{|\Psi_{\vec{k}}|^{2}}_{\text{FS}}-|\braket{\Psi_{\vec{k}}}_{\text{FS}}|^{2}}{2\braket{|\Psi_{\vec{k}}|^{2}}_{\text{FS}}};\label{ZetaSingletForm}
\end{equation}
here $\braket{\dots}_{\text{FS}}$ denotes the average over all momenta
$\vec{k}$ on the Fermi surfaces of the system (normalized such that
$\braket{1}_{\text{FS}}=1$). Note that our assumption of disorder
neglects the fact that the wavefunctions at the Fermi surfaces are
composed of $\vec{k}$-dependent superpositions of spin and different
orbitals, which is expected \cite{DisorderSOCFu,TimmonsPdTe2,OurDisorderSOC,BrydonScattering}
to reduce the impact of disorder on $T_{c}$ further. Therefore, the
following values of $\zeta$ should technically be viewed as upper
bounds.

It holds $\zeta=0$ in \equref{ZetaSingletForm} if $\Psi_{\vec{k}}$
is independent of $\vec{k}$, recovering the well-known Anderson theorem
\cite{Anderson}. Therefore, to obtain finite $\zeta$ in \equref{ZetaSingletForm}
for the $A_{1g}$ singlet, we need to allow for momentum dependent
$\Psi_{\vec{k}}$. Let us first assume that this momentum dependence
arises from $\Psi_{\vec{k}}$ varying \textit{within} a closed Fermi
sheet. To illustrate the consequences for $\zeta$, we will for concreteness
focus on a single Fermi surface enclosing the $\Gamma$ point. Let
us approximate it to be spherical, and only include the lowest-order
lattice harmonic ($g$-wave in this case) correction to $\Psi_{\vec{k}}=\Psi_{0}$
that transforms under the trivial IR $A_{1g}$ of $O_{h}$, 
\begin{equation}
\Psi_{\vec{k}}=\Psi_{0}\left(1+\delta[k_{x}^{4}+k_{y}^{4}+k_{z}^{4}]\right).\label{gwavegapanisotropy}
\end{equation}
Here the parameter $\delta$ determines the strength of the momentum-dependent
perturbation and has to be real as a gauge has to exist where $\Psi_{\vec{k}}\in\mathbb{R}$
(due to time-reversal symmetry in the normal state). Note that the
superconductor will be nodal if and only if $-3<\delta<-1$. From
\equref{ZetaSingletForm}, it is straightforward to evaluate $\zeta$
which is found to be 
\begin{equation}
\zeta(\delta)=\frac{8\,\delta^{2}}{5\delta(41\delta+126)+525}.\label{FormOfSensitivityParameter}
\end{equation}
As expected, we have $\zeta(\delta=0)=0$ since the order parameter
is momentum independent when $\delta=0$. The maximal value of $1/2$
is reached when $\delta=-5/3$ for which the Fermi surface average
of $\Psi_{\vec{k}}$ vanishes. For large $|\delta|$, the order parameter
approaches that of the sub-leading, $g$-wave basis function, associated
with a value $\lim_{|\delta|\rightarrow\infty}\zeta(\delta)=8/205\approx0.039$.

\begin{figure}[tb]
\centering
\includegraphics[width=0.9\linewidth]{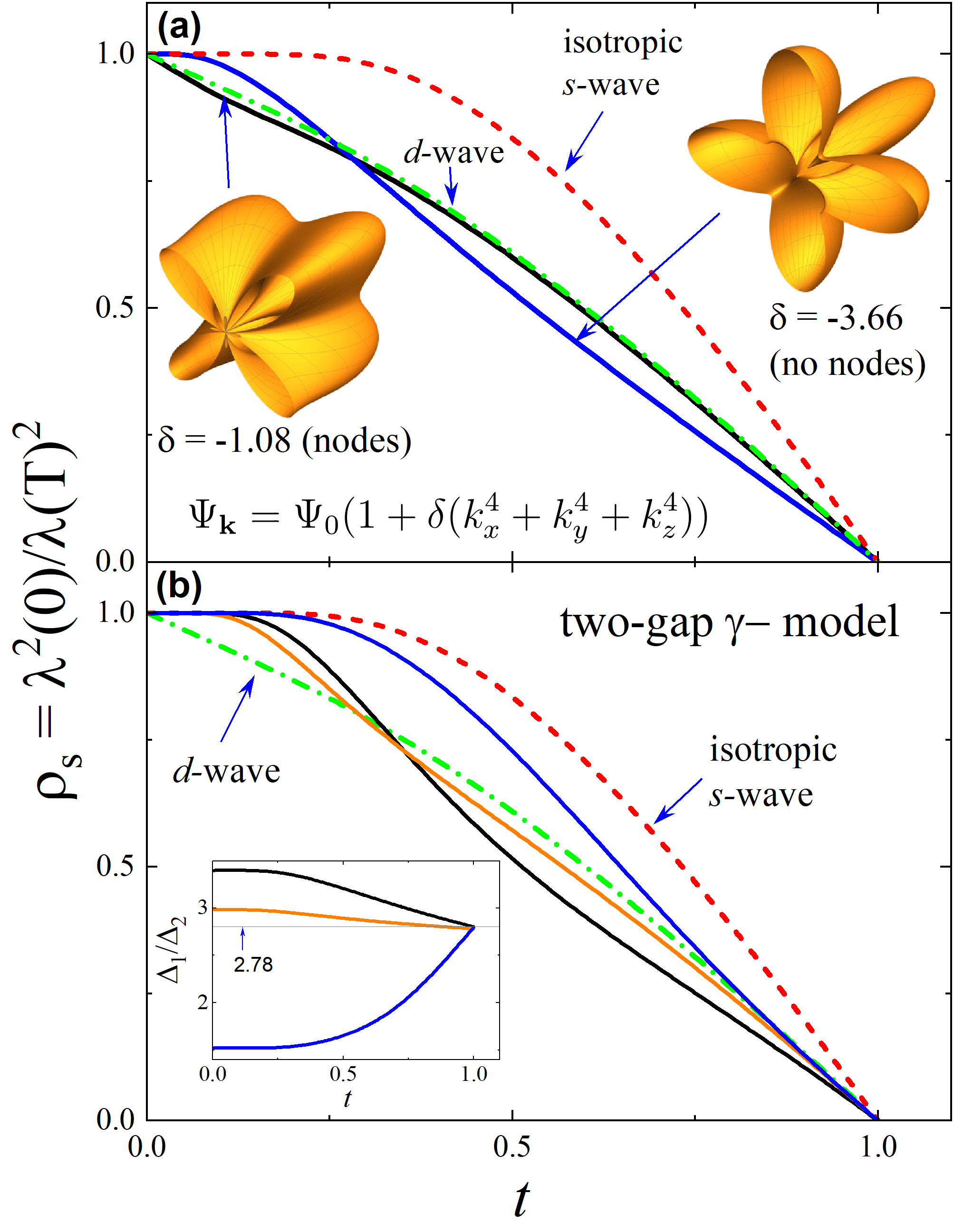} 
\caption{Calculated normalized superfluid density to examine the influence
of gap anisotropy. (a) $g$-wave correction to $s$-wave pairing,
as defined in \equref{gwavegapanisotropy}, with relative strengths
$\delta=-3.66$ (blue) and $-1.08$ (black), needed to reproduce the
observed suppression of $T_{c}$. The 3D structure of the gap in the
reciprocal space is shown as insets. The standard $s$-wave, $\Psi_{\vec{k}}=\Psi_{0}$,
and $d$-wave, $\Psi_{\vec{k}}=\Psi_{0}(k_{x}^{2}-k_{y}^{2})$, cases
are shown by dashed lines. Clearly, the computed superfluid density
is far from the experimental data shown in Fig.~\ref{fig:lambda}.
(b) Two-band $A_{1g}^{++}$ superconductor with two different gap
magnitudes, $\Delta_{1}$ and $\Delta_{2}$. To reproduce the observed
$T_{c}$ suppression, the gap ratio should be $\Delta_{1}/\Delta_{2}=2.78$,
see text. There are many sets of the interaction matrix elements to
obtain that value at $T_{c}$ but with different temperature dependencies
of $\Delta_{1}/\Delta_{2}$ below $T_{c}$ (see inset). We show the
computed $\rho_{s}(T)$ for several choices, but none of them is consistent
with the experimental data. See \appref{SuperfluidStiffnessComp}
for more details of the computations.}
\label{fig:ComputedPenetrationDepth} 
\end{figure}

Most importantly, for the experimental value $\zeta=1/9$, \equref{FormOfSensitivityParameter}
is only consistent with two possible values of $\delta$: either $\delta\approx-1.08$,
which leads to a superconductor with nodal lines, or $\delta\approx-3.66$,
for which the superconductor almost exhibits nodal lines; the associated
anisotropy is quite large, $\Delta_{\text{max}}/\Delta_{\text{min}}\approx12$.
For both values of $\delta$, we have computed the temperature dependence
of the superfluid density $\rho_{s}$, see Fig.~\ref{fig:ComputedPenetrationDepth}(a)
for the results and \appref{SuperfluidStiffnessComp} for more details.
As can be clearly seen, the strong anisotropy or presence of nodes
leads to a $\rho_{s}(T)$ profile that differs significantly from
the observed $s$-wave-like behavior and more closely resembles that
of a $d$-wave state. 
Since none of these two values of $\delta$ are consistent with our
data, we conclude that the momentum dependence of $\Psi_{\vec{k}}$
\textit{on} one (or several) Fermi sheets is not a possible cause
of the observed suppression of $T_{c}$.

\begin{figure}[tb]
\centering
\includegraphics[width=0.9\linewidth]{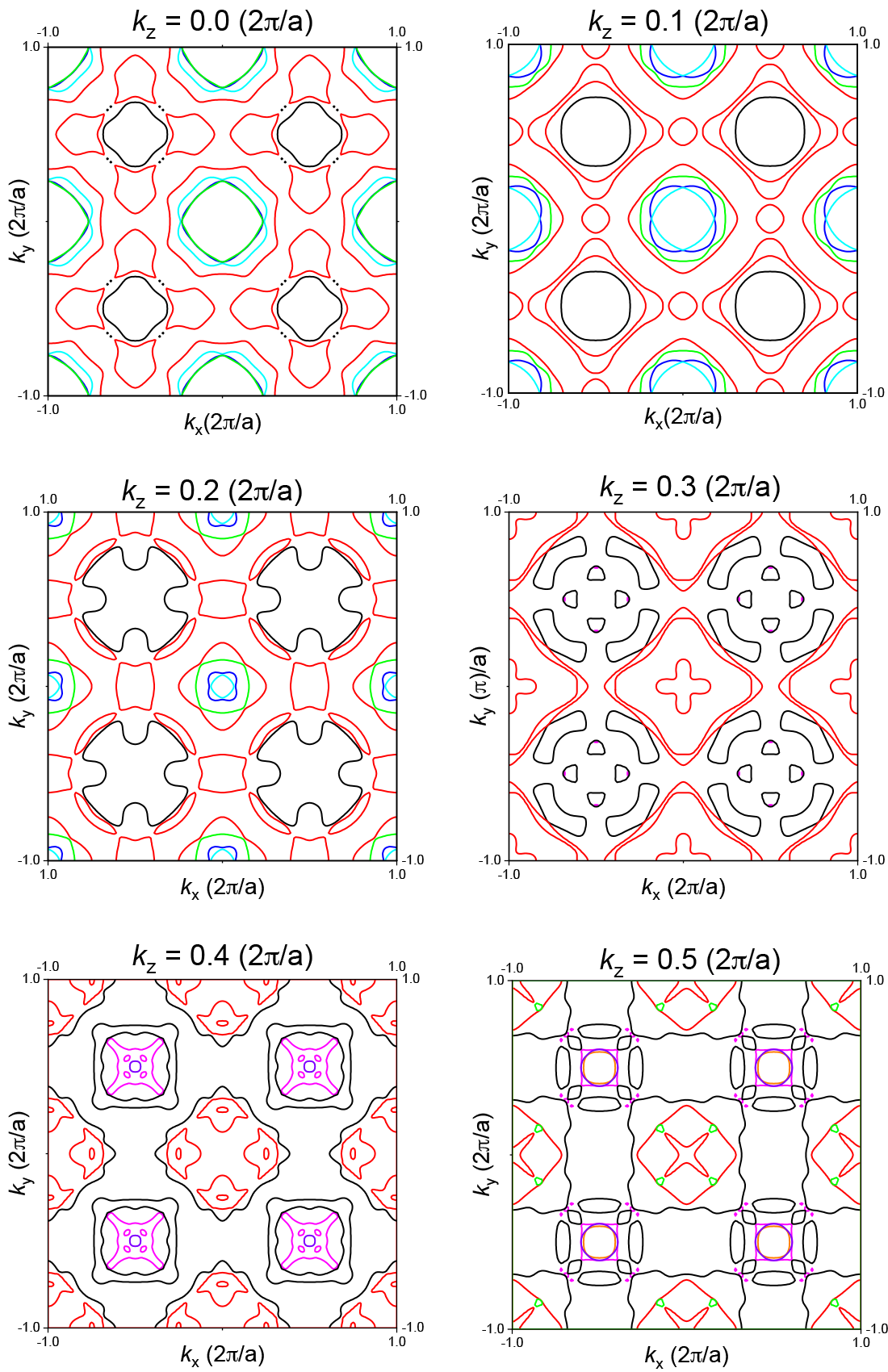} 
\caption{Fermi surface contours at various $k_{z}$ in $\text{Ca}_{3}\text{Rh}_{4}\text{Sn}_{13}$ obtained within
density functional theory (DFT). Different colors denote different
bands crossing the Fermi energy, (see \appref{AppendixDFT} for details).}
\label{fig:FermiSurfaces} 
\end{figure}

Next, we consider the possibility that the order parameter of the
$A_{1g}$ state varies \textit{between different} sheets. 
As can be seen in Fig.~\ref{fig:FermiSurfaces}, the normal state of
$\text{Ca}_{3}\text{Rh}_{4}\text{Sn}_{13}$ has eight bands crossing the Fermi level, giving rise to
very complex Fermi surfaces. Assuming that $\Psi_{\vec{k}}$ is constant
on each Fermi sheet, we write $\Psi_{\vec{k}}=\Delta_{n}$ if $\vec{k}$
belongs to the $n$th sheet. Denoting the density of states at the
Fermi level of the $n$th Fermi surface by $\rho_{n}$, we find 
\begin{equation}
\zeta=\frac{1}{2}\left[1-\frac{\left|\sum_{n}\rho_{n}\Delta_{n}\right|^{2}}{\left(\sum_{n}\rho_{n}|\Delta_{n}|^{2}\right)\sum_{n}\rho_{n}}\right]\label{SecondExpressionForZeta}
\end{equation}
from \equref{ZetaSingletForm}. We note in passing that it is also
possible that the order parameter on different, symmetry unrelated
pockets exhibits non-trivial complex phases, due to frustration \cite{ChubukovSPIS,LaNi},
$\Delta_{n}^{*}\Delta_{n'}^{\phantom{*}}\notin\mathbb{R}$. However,
this can only happen via two (or more) consecutive superconducting
transitions, as a result of time-reversal symmetry. As there are no
indications of multiple transitions in $\text{Ca}_{3}\text{Rh}_{4}\text{Sn}_{13}$ we will assume $\Delta_{n}\in\mathbb{R}$.

Since $\Delta_{n}^{*}\Delta_{n'}^{\phantom{*}}<0$ is expected to
be impossible for a conventional phonon-mediated pairing mechanism
\cite{Brydon,ScheurerPRB2016}, we first focus on the case where $\Delta_{n}^{*}\Delta_{n'}^{\phantom{*}}>0$
for all $n,n'$, which we refer to as the $A_{1g}^{++}$ state. In
the simplest case of only two different gap magnitudes in \equref{SecondExpressionForZeta},
it is straightforward to show via optimization of the respective density
of states that the maximum possible $\zeta$ for given $\Delta_{1}/\Delta_{2}$
reads as 
\begin{equation}
\zeta_{\text{max}}\left(\frac{\Delta_{2}}{\Delta_{1}}\right)=\zeta_{\text{max}}\left(\frac{\Delta_{1}}{\Delta_{2}}\right)=\frac{(\Delta_{2}/\Delta_{1}-1)^{2}}{2(1+\Delta_{2}/\Delta_{1})^{2}}.\label{MaximumExpressionForZeta}
\end{equation}
From this, it is easy to see that $\zeta=1/9$ can only be reached
when $\Delta_{2}/\Delta_{1}>2.78$ (or $\Delta_{1}/\Delta_{2}<0.35$),
which is not consistent with the penetration depth measurement, see
Fig.~\ref{fig:ComputedPenetrationDepth}(b). For reference, $\zeta_{\text{max}}\approx0.004\ll1/9$
assuming a maximal imbalance of $20\%$, $\Delta_{2}/\Delta_{1}=1.2$.
In \appref{DifferentGapsOnFermiSurfaces}, we show that this conclusion
is not altered by allowing for three or more independent gaps. We
also use values of $\rho_{n}$ determined from first-principle calculations
to show that, irrespective of how two different gap magnitudes are
distributed among the various Fermi surface sheets, the minimal gap
anisotropy consistent with $\zeta=1/9$ is $\Delta_{\text{max}}/\Delta_{\text{min}}\approx2.86$.

Since our data cannot be explained by the $A_{1g}^{++}$, we now allow
for $\Delta_{n}^{*}\Delta_{n'}^{\phantom{*}}<0$. Such a multiband $s^{+-}$ state, which we denote by $A_{1g}^{+-}$, cannot be stabilized by
electron-phonon coupling alone and, hence, requires an unconventional
pairing mechanism. As a consequence of the sign change, two different
gap values $\Delta_{1}$ and $\Delta_{2}$ with $\Delta_{1}^{*}\Delta_{2}^{\phantom{*}}<0$
are sufficient to cause much larger $\zeta$ in \equref{SecondExpressionForZeta}:
the maximum possible value of $\zeta=1/2$ is reached when $\rho_{1}|\Delta_{1}|=\rho_{2}|\Delta_{2}|$.
Consequently, for the $A_{1g}^{+-}$ states, the crucial question
is whether $\zeta=1/9$ is too small. In \appref{DifferentGapsOnFermiSurfaces},
we show that there are multiple different ways of distributing $\Delta_{1}$
and $\Delta_{2}$ with $\Delta_{1}/\Delta_{2}\approx-1$, i.e., with
almost identical (and isotropic) gaps, among the various Fermi sheets.
Consequently, the unconventional $A_{1g}^{+-}$ state is thus far
the only option consistent with our measurements.

Finally, let us look into the A$_{1u}$ triplet. As readily follows
from the general expression for $\zeta$ derived in \cite{TimmonsPdTe2},
it holds $\zeta=1/2$ for the A$_{1u}$ triplet state. In fact, $\zeta=1/2$
holds for any other unconventional pairing state such as the $E_{g}$
and $T_{2g}$ singlets in \tableref{PossiblePairingStates}; as already
discussed, these latter two, are less natural candiates for $\text{Ca}_{3}\text{Rh}_{4}\text{Sn}_{13}$
since their gap function is expected to have an anisotropy of about
$2$, while the gap of the A$_{1u}$ triplet state can be completely
isotropic. The value of $\zeta=1/2$ is still too large by about a factor
of four. However, as alluded to above, assuming impurities that have
$\vec{k}$-independent, pseudospin trivial matrix elements on the
Fermi surfaces may not be such a good approximation in a complex multi-orbital
material such as $\text{Ca}_{3}\text{Rh}_{4}\text{Sn}_{13}$. In particular, the presence of spin-orbit
coupling can further reduce $\zeta$ significantly as discussed in
several previous works \cite{TimmonsPdTe2,DisorderSOCFu,OurDisorderSOC,BrydonScattering}.
Therefore, the A$_{1u}$ triplet cannot be excluded based on our observations,
but requires the additional, yet not implausible, assumption of spin-orbit-coupling-induced
suppression of disorder matrix elements between relevant states at
the Fermi surface (see, e.g., \refcite{TimmonsPdTe2} for a general
discussion of this aspect). One observation that further disfavors
the A$_{1u}$ triplet, however, is that the amount of spin-orbit coupling
in the Bloch states at the Fermi surface should vary significantly
among the four stannides studied and yet the suppression of $T_{c}$
with disorder is of the same order in $\text{Ca}_{3}\text{Rh}_{4}\text{Sn}_{13}$ and $\text{Ca}_{3}\text{Ir}_{4}\text{Sn}_{13}$ after
CDW order has been fully suppressed in the latter, see Fig.~\ref{fig:SummaryFigure},
right panel. Taken together, the A$_{1u}$ triplet cannot be rigorously excluded
based on our data but requires more fine-tuning and additional assumptions
than the $A_{1g}^{+-}$ superconductor.

\section{Conclusions\label{Sec:Conclusion}}

We have studied the impact of controlling the number of non-magnetic
defects on the transition temperatures of the superconducting, $T_{c}$,
and CDW, $T_{\text{CDW}}$, phases in the four stoichiometric 3-4-13
stannides $\text{Sr}_{3}\text{Ir}_{4}\text{Sn}_{13}$, $\text{Sr}_{3}\text{Rh}_{4}\text{Sn}_{13}$, $\text{Ca}_{3}\text{Ir}_{4}\text{Sn}_{13}$, and $\text{Ca}_{3}\text{Rh}_{4}\text{Sn}_{13}$. While $T_{\text{CDW}}$
is suppressed with increasing defect concentrations in the three compounds
that exhibit CDW order, the behavior of superconductivity is more
complex, see Fig.~\ref{fig:SummaryFigure}, and reveals non-trivial
microscopic physics. The suppression of $T_{c}$ with weak disorder
is by far the strongest in $\text{Ca}_{3}\text{Rh}_{4}\text{Sn}_{13}$, which does not exhibit any long-range
CDW. Furthermore, $T_{c}$ increases with weak disorder in $\text{Sr}_{3}\text{Rh}_{4}\text{Sn}_{13}$.
All of these findings are consistent with a picture where CDW and
superconductivity compete. 

Quantitatively, the suppression of $T_{c}$ with disorder in $\text{Ca}_{3}\text{Rh}_{4}\text{Sn}_{13}$
is about only $4.5$ times weaker than the theoretical expectation
for a nodal superconducting gap function, such as $d$-wave, with
a vanishing average order parameter on the Fermi surface, see Fig.~\ref{fig:scattering}.
However, the measured temperature dependence of the London penetration
depth, Fig.~\ref{fig:lambda}, indicates a full isotropic gap. Based
on the symmetries of the normal state, we classified the pairing states
in $\text{Ca}_{3}\text{Rh}_{4}\text{Sn}_{13}$ and list those which can have a full gap in \tableref{PossiblePairingStates}.
Among those, only the $A_{1g}$ singlet and $A_{1u}$ triplet are
naturally consistent with the nearly isotropic gap. Based on a quantitative
comparison \cite{TimmonsPdTe2} of theory and the measured disorder-induced
change of $T_{c}$ in $\text{Ca}_{3}\text{Rh}_{4}\text{Sn}_{13}$, a conventional $A_{1g}$ singlet,
where the sign of the order parameter is the same on all Fermi surfaces,
is not consistent with the data. Instead, the $A_{1g}^{+-}$ singlet, a multiband $s^{+-}$ state,
where the sign of the order parameter is different on one (or a small
subset) of the smaller Fermi surfaces, naturally reproduces the observed
suppression of $T_{c}$. While we cannot rigorously exclude the $A_{1u}$
triplet, further assumptions about the matrix elements of the disorder
potential on the Fermi surfaces are required to reduce the impact
of disorder on its critical temperature. In either case, the pairing
mechanism giving rise to the $A_{1g}^{+-}$ or $A_{1u}$ superconductor
cannot \cite{ScheurerPRB2016,Brydon} be based entirely on electron-phonon
coupling, and thus must be unconventional. Similarly, since regular time-reversal-invariant CDW fluctuations cannot induce unconventional pairing \cite{ScheurerPRB2016}, our work motivates further investigations into a possible microscopic origin of unconventional pairing. 

While this conclusion about unconventional pairing only directly applies
to $\text{Ca}_{3}\text{Rh}_{4}\text{Sn}_{13}$, it is natural to expect that the superconductivity has
a very similar nature in all of the studied stannides. We observed an extremely similar superfluid density in $\text{Ca}_{3}\text{Ir}_{4}\text{Sn}_{13}$ (a separate study to be published), which indicates that it is also a fully gapped superconducting state.  As shown in Fig.~\ref{fig:SummaryFigure}, $T_{c}$ is only weakly suppressed in  $\text{Ca}_{3}\text{Ir}_{4}\text{Sn}_{13}$ when CDW is present, but is suppressed at a similar rate to  $\text{Ca}_{3}\text{Rh}_{4}\text{Sn}_{13}$ as soon as CDW is suppressed. Therefore, it is reasonable to conclude that the underlying $T_c$ suppression rate is the same in $\text{Ca}_{3}\text{Rh}_{4}\text{Sn}_{13}$ as in $\text{Ca}_{3}\text{Ir}_{4}\text{Sn}_{13}$, implying similar unconventional pairing. In this sense, $\text{Ca}_{3}\text{Rh}_{4}\text{Sn}_{13}$ could be the key compound to unravel the microscopic physics of superconductivity in the 3-4-13 series. 

\begin{acknowledgments} 

This Research is supported by the U.S. Department of Energy, Office
of Basic Energy Sciences, Materials Science and Engineering Division
through the Ames Laboratory. The Ames Laboratory is operated for the
U.S. Department of Energy by Iowa State University under Contract
No. DE-AC02-07CH11358. Electron irradiation was conducted at the ``SIRIUS" accelerator facility at \'{E}cole Polytechnique (Palaiseau, France) and was supported by EMIR Network proposal No. 20-5925.
L.K. acknowledges support from the U.S. DOE
Early Career Research Program. P.P.O. acknowledges support from the
Research Corporation for Science Advancement via a Cottrell Scholar
Award. Work at Brookhaven National Laboratory was supported by the
U.S. Department of Energy, Office of Basic Energy Science, Division
of Materials Science and Engineering, under Contract No.~DE-SC0012704
(materials synthesis).
\end{acknowledgments} 

\appendix

\section{Pairing states and impact of disorder}

In this appendix, we provide more details on the superconducting pairing
states for $\text{Ca}_{3}\text{Rh}_{4}\text{Sn}_{13}$ and their sensitivity to point-like disorder.

\subsection{Admixture of secondary order parameter}

\label{Admixture} To understand why the superconducting order parameters
in the last two lines of \tableref{PossiblePairingStates} allow
for admixture of a secondary order parameter that can be fully gapped,
let us first focus on the $e_{g}(0,1)$ state. When the order parameter
$\Delta_{\vec{k}}=\eta_{2}^{E_{g}}(2Z_{\vec{k}}^{2}-X_{\vec{k}}^{2}-Y_{\vec{k}}^{2})i\sigma_{y}$
becomes non-zero at $T_{c}$, it reduces the point symmetries not
only in charged but also in charge-$0$ observables, such as the spectrum
$E_{\vec{k}}$ of the Bogoliubov quasi-particles. Formally, this means
that for some $g\in O_{h}$, no $\varphi_{g}\in\mathbb{R}$ exists
such that 
\begin{equation}
\Delta_{g\vec{k}}=e^{i\varphi_{g}}\Delta_{\vec{k}},\quad\forall\,\vec{k}.\label{ConditionForProjSym}
\end{equation}
For the $e_{g}(0,1)$ state, the maximal set of $g\in O_{h}$ for
which a $\varphi_{g}$ in \equref{ConditionForProjSym} exists forms
the subgroup $D_{4h}$ of $O_{h}$; therefore, $E_{\vec{k}}$ will
only be invariant under these symmetries, while $E_{g\vec{k}}\neq E_{\vec{k}}$
for $g\in O_{h}\setminus D_{4h}$ (such as three-fold rotational symmetry).
Since the order parameter of the $e_{g}(0,1)$ state transforms under
the trivial representation, $A_{1g}$, of $D_{4h}$ {[}$\Leftrightarrow\varphi_{g}=0$
in \equref{ConditionForProjSym} for all $g\in D_{4h}${]}, it can
couple linearly to the $A_{1g}$ singlet in \tableref{PossiblePairingStates}.
This coupling requires $O_{h}$ to be broken due to $\eta_{2}^{E_{g}}\neq0$
and is thus a higher-order process in $\eta_{2}^{E_{g}}$. As such,
we expect the admixed component to have a temperature dependence $\propto(T_{c}-T)^{n/2}$,
with $n>1$, close to $T_{c}$. We note that this would be different
in case of the $e_{g}(1,0)$ superconductor with order parameter $\Delta_{\vec{k}}=\eta_{1}^{E_{g}}(X_{\vec{k}}^{2}-Y_{\vec{k}}^{2})i\sigma_{y}$;
while it will also reduce $O_{h}$ to $D_{4h}$, we will have $\varphi_{C_{4}^{z}}=\varphi_{\sigma_{d}}=\pi$
in \equref{ConditionForProjSym} such that the order parameter will
transform as $B_{1g}$ under $D_{4h}$. Being odd under the mirror
planes $\sigma_{d}$ of $D_{4h}$, any $B_{1g}$ singlet will necessarily
have line nodes.

To demonstrate the admixture for $E_{g}$ pairing more explicitly
and determine the exponent $n$ in the temperature dependence of the
secondary order parameter, we will next discuss it on the level of
a Ginzburg-Landau expansion. To this end, we expand the order parameter
in the $E_{g}$ and $A_{1g}$ representation of $O_{h}$ as 
\begin{equation}
\Delta_{\vec{k}}i\sigma_{y}=\eta_{1}^{E_{g}}\sqrt{3}(X_{\vec{k}}^{2}-Y_{\vec{k}}^{2})+\eta_{2}^{E_{g}}(2Z_{\vec{k}}^{2}-X_{\vec{k}}^{2}-Y_{\vec{k}}^{2})+\eta^{A_{1g}}.
\end{equation}
As they transform under different IRs of $O_{h}$, there cannot be
a quadratic coupling of the form $(\eta_{j}^{E_{g}})^{*}\eta^{A_{1g}}$,
but upon noting that $E_{g}\otimes E_{g}\otimes E_{g}=A_{1g}\oplus A_{2g}\oplus3E_{g}$
it is clear that quartic terms of the form $(\eta_{j}^{E_{g}})^{*}(\eta_{k}^{E_{g}})^{*}\eta_{l}^{E_{g}}\eta^{A_{1g}}$
exist. As $\vec{\eta}^{\dagger}\sigma_{x}\vec{\eta}$ and $\vec{\eta}^{\dagger}\sigma_{z}\vec{\eta}$,
with $\vec{\eta}=(\eta_{1}^{E_{g}},\eta_{2}^{E_{g}})^{T}$, transform
as $\sqrt{3}(x^{2}-y^{2})$ and $2z^{2}-x^{2}-y^{2}$ under $O_{h}$,
the following coupling is allowed in the free energy: 
\begin{equation}
\kappa\,(\eta^{A_{1g}})^{*}\left(\vec{\eta}^{\dagger}\sigma_{x}\vec{\eta}\,\eta_{1}^{E_{g}}+\vec{\eta}^{\dagger}\sigma_{z}\vec{\eta}\,\eta_{2}^{E_{g}}\right)+\text{c.c.},\label{CouplingInGL}
\end{equation}
where $\kappa\in\mathbb{R}$ as a consequence of time-reversal symmetry.
In agreement with our discussion above, we find that the coupling
vanishes for the $e_{g}(1,0)$ superconductor, where $\eta_{2}^{E_{g}}=0$;
the same holds for the time-reversal-symmetry-breaking $e_{g}(1,i)$
state for which $\vec{\eta}=(1,\pm i)$. On the other hand, it is
non-zero and given by 
\begin{equation}
-2\kappa|\eta_{2}^{E_{g}}|^{2}\text{Re}[(\eta^{A_{1g}})^{*}\eta_{2}^{E_{g}}]
\end{equation}
for the $e_{g}(0,1)$ pairing phase. We thus see that $\eta_{2}^{E_{g}}\neq0$
will induce a finite $\eta^{A_{1g}}\propto|\eta_{2}^{E_{g}}|^{3}\propto(T_{c}-T)^{3/2}$
close to $T_{c}$ (yielding $n=3$).

While the behavior of $\eta^{A_{1g}}(T)$ and $\eta_{2}^{E_{g}}(T)$
further below $T_{c}$ cannot be captured by the leading-order Ginzburg-Landau
expansion and will depend on microscopic details, we can estimate
the gap anisotropy as a function of the ratio $\eta=\eta^{A_{1g}}/\eta_{2}^{E_{g}}$.
Using, as in the main text, $(X,Y,Z)=(k_{x},k_{y},k_{z})$, the gap
anisotropy of the $e_{g}(0,1)$ state on a spherical Fermi surface
reads as 
\begin{align}
\Delta_{\text{max}}/\Delta_{\text{min}}=\begin{cases}
\frac{2+\eta}{\eta-1} & \eta>1,\\
\infty & \eta\leq1.
\end{cases}
\end{align}
For instance, if we want $\Delta_{\text{max}}/\Delta_{\text{min}}<1.1$,
we need $\eta>31$, i.e., the secondary order parameter has to be
about $30$ times larger than the primary one, which does not seem
to be a natural assumption.

The analysis for the $t_{2g}(1,1,1)$ singlet is similar. In this
case, the coupling analogous to \equref{CouplingInGL} is associated
with the $A_{1g}$ term in $T_{2g}\otimes T_{2g}\otimes T_{2g}=A_{1g}\oplus A_{2g}\oplus2E_{g}\oplus3T_{1g}\oplus4T_{2g}$.

\subsection{General expression for $\zeta$}

\label{ExpressionFoZeta} To be self-contained, we here provide the
general expression for the disorder sensitivity parameter $\zeta$
in \equref{eq:Tcsuppression} derived in \cite{TimmonsPdTe2}.
The central quantity is 
\begin{equation}
C_{\vec{k},\vec{k}'}=\Delta_{\vec{k}}T^{\dagger}W_{\vec{k},\vec{k}'}-t_{W}W_{\vec{k},\vec{k}'}\Delta_{\vec{k}'}T^{\dagger},\label{FormOfTheCommutator}
\end{equation}
which is either a commutator or an anti-commutator depending on whether
we consider time-reversal-even ($t_{W}=+1$) or -odd ($t_{W}=-1$)
disorder, respectively; it also appeared in the generalized Anderson
theorem of \cite{Scheurer2016,Hoyer2015}. In \equref{FormOfTheCommutator},
$\Delta_{\vec{k}}$ is the superconducting order parameter at $T_{c}$,
in our case a $2\times2$ matrix in pseudospin space, and $T$ is
the unitary part of the time-reversal operator ($T=i\sigma_{y}$ for
the states in \tableref{PossiblePairingStates}). Finally, $W_{\vec{k},\vec{k}'}$
are the matrix elements of the impurity potential $W$ with respect
to the Bloch states, $\ket{\vec{k},s}$, at the Fermi surface, i.e.,
$(W_{\vec{k},\vec{k}'})_{s,s'}=\braket{\vec{k},s|W|\vec{k}',s'}$,
with $s$ labeling all bands including spin.

Defining the Fermi-surface Frobenius norm according to $||C||_{\text{F}}^{2}:=\frac{1}{N_{\text{FS}}^{2}}\sum_{\vec{k},\vec{k}'\in\text{FS}}\text{tr}\left[C_{\vec{k},\vec{k}'}^{\dagger}C_{\vec{k},\vec{k}'}^{\pdagger}\right]$,
where $\vec{k}\in\text{FS}$ indicates that the sum involves all momenta
in a finite vicinity around the Fermi surfaces and $N_{\text{FS}}=\sum_{\vec{k}\in\text{FS}}$,
we can write \cite{TimmonsPdTe2} 
\begin{equation}
\zeta=\frac{||C||_{\text{F}}^{2}}{2\,\text{tr}\left[W^{\dagger}W\right]\braket{\text{tr}[\Delta_{\vec{k}}^{\dagger}\Delta_{\vec{k}}^{\pdagger}]}_{\text{FS}}},\label{ZetaExpression}
\end{equation}
where $\braket{f_{\vec{k}}}_{\text{FS}}:=\frac{1}{N_{\text{FS}}}\sum_{\vec{k}\in\text{FS}}f_{\vec{k}}$
denotes the normalized Fermi surface average, as also used in the
main text, see \equref{ZetaSingletForm}.

Due to the generality of \equref{ZetaExpression}, it can be readily
applied in many different systems, see, e.g., \refscite{LaNi,TBGApplication}
for two recent applications. Most importantly for us here, \equref{ZetaSingletForm}
in the main text is readily derived by focusing on $\Delta_{\vec{k}}\in\mathbb{C}^{2\times2}$,
$\vec{k}$-dependent pseudospin-singlet pairing, $\Delta_{\vec{k}}=\Psi_{\vec{k}}i\sigma_{y}$
and scalar non-magnetic ($t_{W}=+1$) disorder of the simple form
$W=W_{\vec{k},\vec{k}'}=W_{0}\sigma_{0}$, $W_{0}\in\mathbb{R}$.

\global\long\def\arraystretch{1}%
 
\begin{table}[tb]
\begin{centering}
\caption{Density of states of the bands per conventional unit cell ($Z=2$) of $\text{Ca}_{3}\text{Rh}_{4}\text{Sn}_{13}$ at the Fermi level ordered
by decreasing magnitude.}
\label{DensityOfStates} 
\par\end{centering}
\begin{ruledtabular}
\begin{tabular}{ccc}
$n$  & band & $\rho_{n}\:\left(\nicefrac{\text{states}}{\text{eV cell}}\right)$ \tabularnewline 
\hline 
1  & \#272  & 13.73 \tabularnewline
2  & \#273  & 9.534 \tabularnewline
3  & \#274  & 0.847 \tabularnewline
4  & \#275  & 0.826 \tabularnewline
5  & \#271  & 0.732 \tabularnewline
6  & \#276  & 0.670\tabularnewline
7  & \#270  & 0.135\tabularnewline
8  & \#269  & 0.105 \tabularnewline
\end{tabular}
\end{ruledtabular}

\end{table}

\subsection{Different gaps on different Fermi sheets}

\label{DifferentGapsOnFermiSurfaces} Finally, we discuss in more
details which order parameter ratios $\Delta_{n}/\Delta_{1}$ in \equref{SecondExpressionForZeta}
are consistent with the observed $\zeta=1/9$.

In our DFT calculations for $\text{Ca}_{3}\text{Rh}_{4}\text{Sn}_{13}$ (with details in \appref{AppendixDFT})
we identify eight bands that give rise to Fermi surfaces, see Fig.~\ref{fig:FermiSurfaces}.
Their respective density of states at the Fermi level, $\rho_{n}$,
in decreasing order of magnitude are listed in \tableref{DensityOfStates}.
In principle, the order parameter can be different for any of these
bands. For simplicity, we will first assume that there are only two
different values, $\Delta_{1}$ and $\Delta_{2}$, and the three bands
with smallest $\rho_{n}$ are combined into one, i.e., we take them
to exhibit the same $\Delta_{n}$; this amounts to studying the effective
six band problem with respective density of states 
\begin{align}
\begin{split}\rho'_{n} & =\rho_{n},\qquad\qquad1\leq n\leq5,\\
\rho'_{6} & =\rho_{6}+\rho_{7}+\rho_{8}.
\end{split}
\label{EffectiveSixBand}
\end{align}
There are still many (31) inequivalent ways of distributing two gaps
on the six Fermi surfaces, as listed in \tableref{PossibleAnisotropies}
together with the associated anisotropy ratio consistent with $\zeta=1/9$.
We can see that the smallest possible anisotropy ratio for the $A_{1g}^{++}$
state is $2.86$. We have checked that this value does not change
when allowing for $\Delta_{1}$ and $\Delta_{2}$ to be distributed
arbitrarily on all eight Fermi surfaces in \tableref{DensityOfStates}.
As it should be, this value is larger than the theoretical lower bound
(for $\zeta=1/9$) of $(11+6\sqrt{2})/7\approx2.78$ based on \equref{MaximumExpressionForZeta}
for arbitrary ratio of the density of states; due to the multitude
of different Fermi surfaces, it is also natural that the $A_{1g}^{++}$
state can almost reach this theoretical bound.

For the $A_{1g}^{+-}$ state, there are several solutions with $|\Delta_{1}|/|\Delta_{2}|$
very close to $1$ already in the six-band model, see \tableref{PossibleAnisotropies}.
As can also be seen in the table, this is possible for distributions
of order parameters where the sign change happens between a set of
Fermi surfaces and its complement exhibiting a ratio of density of
states of about 6-7\%.

\begin{table}[tb]
\begin{centering}
\caption{Ratio of order parameters consistent with $\zeta=1/9$ in \equref{SecondExpressionForZeta}
for all possible independent distributions of the two different values,
$\Delta_{1}$ and $\Delta_{2}$, among the six sets of Fermi sheets
defined in \equref{EffectiveSixBand}. Here $\mathcal{S}$ defines
the set of sheets with order parameter $\Delta_{1}$, while the order
parameter is $\Delta_{2}$ on the complement $\bar{\mathcal{S}}=\{1,2,3,4,5,6\}\setminus\mathcal{S}$.
The relative fraction of the density of states of $\mathcal{S}$ is
denoted by $\nu_{\mathcal{S}}:=\sum_{n\in\mathcal{S}}\rho_{n}/\sum_{n\in\bar{\mathcal{S}}}\rho_{n}$.
For a more clear representation of the gap anisotropy, we define $\Delta_{a}/\Delta_{b}:=\text{max}\{\Delta_{1}/\Delta_{2},\Delta_{2}/\Delta_{1}\}$.}
\label{PossibleAnisotropies} 
\par\end{centering}
\begin{ruledtabular}
\begin{tabular}{cccc}
$\mathcal{S}$  & $\nu_{\mathcal{S}}$  & $(\Delta_{a}/\Delta_{b})_{1}$  & $(\Delta_{a}/\Delta_{b})_{2}$ \tabularnewline
\hline 
\{1\}  & 1.07  & 3.39  & 3.21 \tabularnewline
\{2\}  & 0.56  & 2.86  & 4.91 \tabularnewline
\{3\}  & 0.03  & 4.37  & $-$0.56 \tabularnewline
\{4\}  & 0.03  & 4.41  & $-$0.55 \tabularnewline
\{5\}  & 0.03  & 4.59  & $-$0.50 \tabularnewline
\{6\}  & 0.04  & 4.27  & $-$0.60 \tabularnewline
\{1,2\}  & 7.02  & $-$0.35  & 3.03 \tabularnewline
\{1,3\}  & 1.21  & 3.61  & 3.09 \tabularnewline
\{1,4\}  & 1.21  & 3.61  & 3.09 \tabularnewline
\{1,5\}  & 1.19  & 3.58  & 3.10 \tabularnewline
\{1,6\}  & 1.23  & 3.63  & 3.08 \tabularnewline
\{2,3\}  & 0.64  & 2.92  & 4.30 \tabularnewline
\{2,4\}  & 0.64  & 2.91  & 4.31 \tabularnewline
\{2,5\}  & 0.63  & 2.91  & 4.37 \tabularnewline
\{2,6\}  & 0.65  & 2.92  & 4.26 \tabularnewline
\{3,4\}  & 0.07  & 3.55  & $-$0.93 \tabularnewline
\{3,5\}  & 0.06  & 3.61  & \textbf{$-$0.99} \tabularnewline
\{3,6\}  & 0.07  & 3.51  & $-$0.88 \tabularnewline
\{4,5\}  & 0.06  & 3.63  & \textbf{$-$0.99} \tabularnewline
\{4,6\}  & 0.07  & 3.52  & $-$0.90 \tabularnewline
\{5,6\}  & 0.07  & 3.57  & \textbf{$-$0.95} \tabularnewline
\{1,2,3\}  & 9.77  & $-$0.57  & 3.22 \tabularnewline
\{1,2,4\}  & 9.68  & $-$0.57  & 3.22 \tabularnewline
\{1,2,5\}  & 9.29  & $-$0.54  & 3.19 \tabularnewline
\{1,2,6\}  & 10.05  & $-$0.59  & 3.24 \tabularnewline
\{1,3,4\}  & 1.38  & 3.91  & 2.99 \tabularnewline
\{1,3,5\}  & 1.36  & 3.87  & 3.00 \tabularnewline
\{1,3,6\}  & 1.4  & 3.94  & 2.98 \tabularnewline
\{1,4,5\}  & 1.35  & 3.86  & 3.00 \tabularnewline
\{1,4,6\}  & 1.39  & 3.93  & 2.98 \tabularnewline
\{1,5,6\}  & 1.37  & 3.89  & 2.99 \tabularnewline
\end{tabular}
\end{ruledtabular}

\end{table}

One might wonder whether more than two different values of $\Delta_{n}$
in \equref{SecondExpressionForZeta} will allow for an $A_{1g}^{++}$
state with smaller gap anisotropy, 
\begin{equation}
A_{\Delta}:=\max_{n,n'}\frac{\Delta_{n}}{\Delta_{n'}},
\end{equation}
for given $\zeta$ (in our case $\zeta=1/9$). Instead of systematically
studying all possible ways of distributing $N>2$ different order
parameters, $\Delta_{n}>0$, $n=1,2,\dots N$, on the eight different
Fermi surfaces in \tableref{DensityOfStates}, we here derive a lower
bound on $A_{\Delta}$. To this end, let us assume we are given $\Delta_{n}>0$
which we order, without loss of generality, such that $\Delta_{n}>\Delta_{n+1}$.
It is not difficult to show that the maximum value of $\zeta$ in
\equref{SecondExpressionForZeta} is reached when $\rho_{n}=0$ for
all $n\neq1,N$. Consequently, only the smallest and largest $\Delta_{n}$
enter and we are back to the case with only two gaps, which we have
already analyzed in Section~\ref{ConstraintOnPairingFromDisorder} of
the main text; the maximum value $\zeta_{\text{max}}$ thus only depends
on $\Delta_{1}/\Delta_{N}=A_{\Delta}$ with form given in \equref{MaximumExpressionForZeta},
i.e., 
\begin{equation}
\zeta_{\text{max}}(\{\Delta_{n}\})=\frac{(A_{\Delta}-1)^{2}}{2(1+A_{\Delta})^{2}},
\end{equation}
irrespective of how many different $\Delta_{n}$ are taken into account.
Specifically, the lower bound for $\zeta=1/9$, $A_{\Delta}>(11+6\sqrt{2})/7\approx2.78$,
still applies and the $A_{1g}^{++}$ state with three or more different
gaps is not a natural candidate state either.

\section{Superfluid density in different models}

\label{SuperfluidStiffnessComp} Having established in Section~\ref{ConstraintOnPairingFromDisorder}
which fully gapped conventional singlets are consistent with the observed
suppression of $T_{c}$ with impurity concentration, we next investigate
more quantitatively how the respective temperature dependence of the
penetration depth or superfluid density compares with that measured
experimentally (see Fig.~\ref{fig:lambda}).

\subsection{Anisotropic, single Fermi surface}

We first consider the anisotropic singlet on a single, isotropic Fermi
surface as defined in \equref{gwavegapanisotropy}. As discussed
in the main text, only the values of $\delta=-1.08$ and $\delta=-3.66$
reproduce the observed $T_{c}$ suppression. The former is nodal and
cannot possibly explain the exponential attenuation of the penetration
depth. The latter is not nodal, but highly anisotropic. To see whether
this anisotropy is consistent with the superfluid density $\rho_{s}$
of Fig.~\ref{fig:lambda}, we computed $\rho_{s}(T)$ for this model.

The calculations followed the Eilenberger formalism with a common
ansatz that temperature and angular parts of the order parameter can
be separated, $\Delta\left(T,\vec{k}_{F}\right)=\Psi\left(T\right)\Omega\left(\vec{k}_{F}\right)$,
where $\vec{k}_{F}$ is Fermi wave vector and the angular part obeys
the normalization condition for the Fermi surface average, $\left\langle \Omega^{2}\right\rangle _{\text{FS}}=1$
\cite{Kogan2021}. Specifically, for the anisotropic $A_{1g}$ state
in \equref{gwavegapanisotropy}, the angular part in spherical coordinates,
$\vec{k}_{F}=k_{F}(\sin\theta\cos\varphi,\sin\theta\sin\varphi,\cos\theta)$,
reads as 
\begin{equation}
\Omega\left(\theta,\varphi\right)=\frac{1+\delta\left[(\sin\theta\cos\varphi)^{4}+(\sin\theta\sin\varphi)^{4}+\cos^{4}\theta\right]}{\sqrt{1+(6/5)\delta+(41/105)\delta^{2}}}.
\end{equation}

The temperature-dependent order parameter magnitude, $\Psi\left(T\right)$,
is then obtained by solving the Eilenberger self-consistency equation
and after that any thermodynamic quantity, including superfluid density,
is calculated. The result for both values of $\delta$ is shown in
Fig.~\ref{fig:ComputedPenetrationDepth}(a) along with the curves for
a weak-coupling isotropic $s$-wave BCS ($\Omega=1$) and $d$-wave
($\Omega=\sqrt{2}\cos2\varphi$) order parameters. The inset shows
the angular dependence of the gap magnitude, $|\Omega(\vec{k}_{F})|$,
for the same two values of $\delta$. Clearly, $\rho_{s}(T)$ differs
strongly from $s$-wave behavior and, hence, from the data in Fig.~\ref{fig:lambda}
for all of these models.

\begin{figure}[tb]
\centering %
\begin{tabular}{c}
\includegraphics[clip,width=0.95\linewidth]{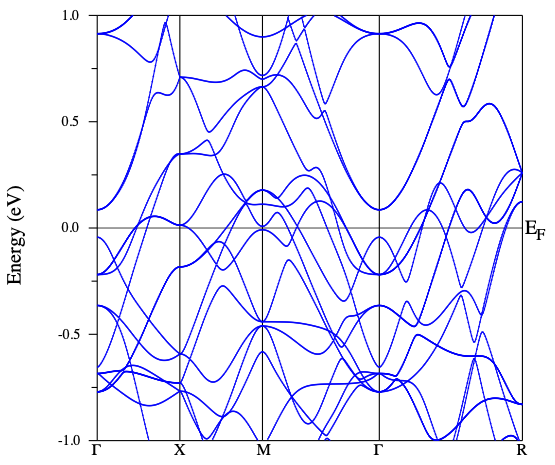} \tabularnewline
\includegraphics[clip,width=0.95\linewidth]{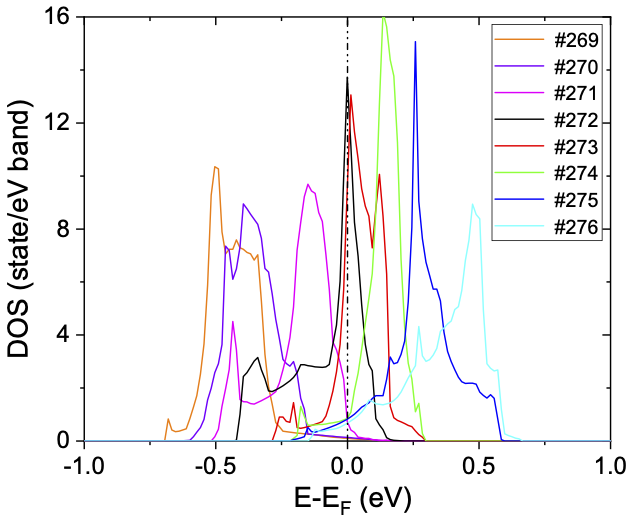} \tabularnewline
\end{tabular}%
\caption{(Top) Band structures and (bottom) partial density of states of the
eight bands across the Fermi level in $\text{Ca}_{3}\text{Rh}_{4}\text{Sn}_{13}$.}
\label{fig:band_dos} 
\end{figure}

\subsection{Isotropic self-consistent two-band model}

Another way to obtain substantial $T_{c}$ suppression in a conventional
superconductor is to consider a two band system with two isotropic
$s$-wave bands of different amplitude but same sign, denoted as $A_{1g}^{++}$
in the main text.

To compute the superfluid density $\rho_{s}$ for this scenario, we
use the self-consistent Eilenberger scheme, called the $\gamma$-model,
which is detailed in \refcite{Kogan2009}. Starting with an interaction
matrix containing two intra-band and one inter-band interaction constants,
a system of $2\times2$ self-consistency equations yields two order
parameters from which the total $\rho_{s}$ can be calculated. Note
that the temperature dependencies of the order parameters no longer
follow the standard isotropic single-band curve, implying that the
gap ratio is temperature-dependent; furthermore, its precise temperature
evolution depends crucially on the interaction parameters while the
amount of $T_{c}$ suppression is dictated by the gap ratio at $T_{c}$
(see \appref{ExpressionFoZeta}). Therefore, we selected several
combinations of the interaction parameters, varying intra- and inter-band
contributions, with the constraint that the gap ratio at $T_{c}$
is $\Delta_{1}/\Delta_{2}=2.78$, needed to obtain the measured $T_{c}$
suppression. In Fig.~\ref{fig:ComputedPenetrationDepth}(b), we present
the resulting temperature dependence of $\rho_{s}$ (main panel) and
of the gap ratio $\Delta_{1}/\Delta_{2}$ (inset) for three different
sets of interaction parameters, with roughly constant, increasing,
and decreasing $\Delta_{1}/\Delta_{2}$. While the low-temperature
behavior exhibits saturation, it occurs below the temperature where
the small gap saturates, much lower that $T_{c}/3$ of isotropic s-wave.
Most importantly, as before, none of these models of conventional
pairing agree with the measure superfluid density. We therefore have
a strong case in favor of unconventional pairing.

\textcolor{white}{.}

\section{Density functional theory calculations\label{AppendixDFT}}

We carry out first principles calculations to
investigate the electronic structures in $\text{Ca}_{3}\text{Rh}_{4}\text{Sn}_{13}$. $\text{Ca}_{3}\text{Rh}_{4}\text{Sn}_{13}$ crystallizes
in the cubic Yb$_{3}$Rh$_{4}$Sn$_{13}$-type ($Pm\overline{3}n$,
space group no.~223) structure. The primitive cell contains two formula
units. Ca atoms occupy the $6c\,(\overline{4}m2)$ site, and Rh atoms
occupy the $8e\,(32)$ site. The Sn atoms are divided into two sublattices;
out of 13 Sn atoms in one formula unit, 12 Sn$_{1}$ atoms occupy
the $24k\,(m)$ site, and one Sn$_{2}$ atom occupies the $2a\,(m\overline{3})$
site. We adopt the experimental crystal structure parameters~\cite{hodeau1982ssc}
in all calculations.

Density functional theory (DFT) calculations are performed using a
full-potential linear augmented plane wave (FP-LAPW) method, as implemented
in \textsc{wien2k}~\cite{WIEN2k}. The generalized gradient approximation
of Perdew, Burke, and Ernzerhof~\cite{perdew1996prl} is used for
the exchange-correlation potentials. To generate the self-consistent
potential and charge, we employed $R_{\text{MT}}\cdot K_{\text{max}}=8.0$
with Muffin-tin radii $R_{\text{MT}}=$ 2.2, 2.4, and 2.4\,{a.u.},
for Ca, Rh, and Sn, respectively. The $k$-point integration is performed
using a tetrahedron method with Blöchl corrections~\cite{blochl1994prb}
with 119 $k$-points in the irreducible Brillouin zone (BZ). The calculations
are iterated until the charge difference between consecutive iterations
is smaller than $10^{-4}$ $e$ and the total energy difference is
lower than 0.01~mRy.

Figure~\ref{fig:band_dos} shows the DFT band structure along the
$\Gamma$--$X$--$M$--$\Gamma$--$R$ high-symmetry path and
band-resolved partial density of states (PDOS) near the Fermi level.
There are eight bands across the Fermi level. Figure~\ref{fig:FermiSurfaces}
shows the Fermi surface contours calculated at various $k_{z}$. We
use the same color scheme to denote the eight bands in the Fermi surface
contours (Fig.~\ref{fig:FermiSurfaces}) and the band-resolved PDOS
(Fig.~\ref{fig:band_dos}(bottom)) plots.


\begin{thebibliography}{99}

\bibitem{Mathur} N. D. Mathur, F.M. Grosche, S. R. Julian, I. R.
Walker, D., Monthoux M. Freye, R. K. W. Haselwimmer, and G. G. Lonzarich,
 Nature (London) \textbf{394}, 39 (1998).

\bibitem{Belitz2005}D. Belitz and T. Vojta, Rev. Mod. Phys. \textbf{77}, 2 (2005).

\bibitem{Monthoux} P.~Monthoux, D.~Pines, and G.~G.~Lonzarich,
 Nature (London) \textbf{433}, 226 (2005).

\bibitem{Keimer} B.~Keimer, S.~A.~Kivelson, M.~R.~Norman, S.~Uchida,
and J.~Zaanen, Nature (London) \textbf{518}, 179 (2015).

\bibitem{Norman2011}M. R. Norman, Science \textbf{332}, 6026 (2011).

\bibitem{Levchenko2013}A. Levchenko, M. G. Vavilov, M. Khodas, and A. V. Chubukov, Phys. Rev. Lett. \textbf{110}, 17 (2013).

\bibitem{CPLT} C. Proust and L. Taillefer, Annu. Rev. Condens. Matt. Phys. \textbf{10}, 409 (2019).

\bibitem{Khodas2020}M. Khodas, M. Dzero, and A. Levchenko, Phys.
Rev. B \textbf{102}, 184505 (2020). 

\bibitem{Chubukov2020}A. V. Chubukov, A. Abanov, Y. Wang, and Y.-M. Wu, Ann. Phys. \textbf{417}, 168142 (2020).

\bibitem{Hashimoto2012}K. Hashimoto, K. Cho, T. Shibauchi, S. Kasahara,
Y. Mizukami, R. Katsumata, Y. Tsuruhara, T. Terashima, H. Ikeda, M.
A. Tanatar, H. Kitano, N. Salovich, R. W. Giannetta, P. Walmsley,
A. Carrington, R. Prozorov, and Y. Matsuda, Science \textbf{336},
1554-1557 (2012). 

\bibitem{Wang2018}C. G. Wang, Z. Li, J. Yang, L. Y. Xing, G. Y. Dai, X. C. Wang, C. Q. Jin, R. Zhou, and G.-q. Zheng, Phys. Rev. Lett. \textbf{121}, 16 (2018).

\bibitem{Joshi2020}K. R. Joshi, N. M. Nusran, M. A. Tanatar, K. Cho,
S. L. Bud'ko, P. C. Canfield, R. M. Fernandes, A. Levchenko, and R.
Prozorov, New J. Phys. \textbf{22}, 053037 (2020). 

\bibitem{Cheng2015}J.-G. Cheng, K. Matsubayashi, W. Wu, and J., Phys. Rev. Lett. \textbf{114}, 11 (2015).

\bibitem{Park2011}T. Park, V. A. Sidorov, H. Lee, F. Ronning, E. D. Bauer, J. L. Sarrao, and J. D. Thompson, J. Phys: Cond. Matt. \textbf{23}, 9 (2011).

\bibitem{Custers} J. Custers, P. Gegenwart, H. Wilhelm, K. Neumaier,
Y. Tokiwa, O. Trovarelli, C. Geibel, F. Steglich, C. P\'{E}pin, and P.
Coleman, Nature \textbf{424}, 524 (2003).

\bibitem{Paglione} Johnpierre Paglione, M. A. Tanatar, D. G. Hawthorn, Etienne Boaknin, R. W. Hill, F. Ronning, M. Sutherland, Louis Taillefer, C. Petrovic, and P. C. Canfield, Phys. Rev. Lett. \textbf{91}, 246405 (2003).
 

\bibitem{Budko} S. L. Bud'ko, E. Morosan, and P. C. Canfield, Phys. Rev. B \textbf{69}, 014415 (2004).


\bibitem{Mutka} H. Mutka, Phys. Rev. B \textbf{28}, 2855 (1983).



\bibitem{Shibauchidisorder} Yuta Mizukami, Marcin Ko\'{n}czykowski,
Kohei Matsuura, Tatsuya Watashige, Shigeru Kasahara, Yuji Matsuda,
and Takasada Shibauchi, J. Phys. Soc. Jpn. \textbf{86}, 083706 (2017).

\bibitem{NbSe2disorder} Kyuil Cho, M. Ko\'{n}czykowski, S. Teknowijoyo,
M. A. Tanatar, J. Guss, P. B. Gartin, J. M. Wilde, A. Kreyssig, R.
J. McQueeney, A. I. Goldman, V. Mishra, P. J. Hirschfeld and R. Prozorov, Nat. Comm. \textbf{9}, 2796 (2018).




\bibitem{Abanov2003}A. Abanov, A. V. Chubukov, and J. Schmalian, Adv. Phys. \textbf{52}, 3 (2003).


\bibitem{QCPhf} P.~Coleman and A.~J.~Schofield, Nature \textbf{433 }, 226 (2005).
 
\bibitem{CB} P.~ C. Canfield and S.~L. Bud'ko, Annu. Rev. Condens. Matt. Phys. \textbf{1}, 27 (2010).
 
\bibitem{DCJ2010} D.~C.~Johnston, Adv. Phys. \textbf{59},~803 (2010).

\bibitem{paglione2010} J.~Paglione and R. L. Greene, Nat. Phys. \textbf{6},~645 (2010).
 
\bibitem{carrington} T. Shibauchi, A. Carrington, and Y. Matsuda,
Annu. Rev. Condens. Matter Phys. \textbf{5}, 113 (2014).

\bibitem{Shibauchireview} Takasada Shibauchi, Tetsuo Hanaguri, and
Yuji Matsuda, J. Phys. Soc. Jpn. \textbf{89}, 102002 (2020).

\bibitem{Eulers} Felix Eilers, Kai Grube, Diego A. Zocco, Thomas
Wolf, Michael Merz, Peter Schweiss, Rolf Heid, Robert Eder, Rong Yu,
Jian-Xin Zhu, Qimiao Si, Takasada Shibauchi, and Hilbert v. L\:ohneysen
Phys. Rev. Lett. \textbf{116}, 237003 (2016).

\bibitem{JHChu} Paul Malinowski, Qianni Jiang, Joshua J. Sanchez,
Joshua Mutch, Zhaoyu Liu, Preston Went, Jian Liu, Philip J. Ryan, 
Jong-Woo Kim \& Jiun-Haw Chu, Nat. Phys. \textbf{16}, 1189 (2020).

\bibitem{Klintberg2012} Lina E. Klintberg, Swee K. Goh, Patricia
L. Alireza, Paul J. Saines, David A. Tompsett, Peter W. Logg, Jinhu
Yang, Bin Chen, Kazuyoshi Yoshimura, and F. Malte Grosche, Phys. Rev. Lett. \textbf{109}, 237008 (2012).

\bibitem{Goh2015} S. K. Goh, D. A. Tompsett, P. J. Saines, H. C.
Chang, T. Matsumoto, M. Imai, K. Yoshimura, and F. M. Grosche, Phys.
Rev. Lett. \textbf{114}, 097002 (2015).

\bibitem{Veiga2020} L. S. I. Veiga, J. R. L. Mardegan, M. v. Zimmermann,
D. T. Maimone, F. B. Carneiro, M. B. Fontes, J. Strempfer, E. Granado,
P. G. Pagliuso, and E. M. Bittar, Phys. Rev. B \textbf{101}, 104511 (2020).

\bibitem{Peierls}R. E. Peierls, \textit{Quantum Theory of Solids}, (Oxford University Press, Oxford, U.K., 2001).

\bibitem{CDW} J. A. Wilson, F. Di Salvo, and S. Mahajan, Adv. Phys. \textbf{24}, 117 (1975).


\bibitem{Pouget2016}J. P. Pouget, Comptes Rendus Phys. \textbf{17}, 332–356 (2016).

\bibitem{Jerome} K.~Bechgaard and D.~Jerome, Sci. Am. \textbf{247}, 52 (1982).

\bibitem{NbSe2nesting} F. Weber, R. Hott, R. Heid, L. L. Lev, M.
Caputo, T. Schmitt, and V. N. Strocov, Phys. Rev. B \textbf{97}, 235122 (2018).
 
\bibitem{TaSe2nesting} Y. W. Li, J. Jiang, H. F. Yang, D. Prabhakaran,
Z. K. Liu, L. X. Yang, and Y. L. Chen, Phys. Rev. B \textbf{97}, 115118 (2018).

\bibitem{TaS2nesting} I. Guillamon, H. Suderow, J.~G. Rodrigo, S.~Vieira,
P. Rodi\'{E}re, L. Cario, E. Navarro-Moratalla, C. Marta-Gastaldo and
E. Coronado, New J. Phys. \textbf{13}, 103020 (2011).



\bibitem{Remeika} J. P. Remeika, G.~P.~Espinosa, A~.S.~Cooper,
H~.Barz, J~.~M.Rowell, D.~B.~McWhan, J.~M.~Vandenberg, D.~E.~Moncton, Z.~Fisk, L.~D.~Woolf, H.~C.~Hamaker, M.~B.~Maple, G.~Shirane, and W.Thomlinson, Solid State Comm. \textbf{34}, 923 (1980).
 
 \bibitem{Mazzone2015}D. G. Mazzone, S. Gerber, J. L. Gavilano, R. Sibille, M. Medarde, B. Delley, M. Ramakrishnan, M. Neugebauer, L. P. Regnault, D. Chernyshov, A. Piovano, T. M. Fern\'{a}ndez-D\'{i}az, L. Keller, A. Cervellino, E. Pomjakushina, K. Conder, and M. Kenzelmann,  \prb \textbf{92}, 024101 (2015).

 
\bibitem{Tompsett2014} D.~A.~Tompsett, Phys. Rev. B {\bf 89}, 075117 (2014).
 
 
\bibitem{Dalton2021}
Manuel Feig,   Lev Akselrud,   Mykhaylo Motylenko,   Matej Bobnar,   J\"og Wagler, Kristina O. Kvashnina,   Volodymyr Levytskyi, David Rafaja, Andreas Leithe-Jasper,   and  Roman Gumeniuk, Dalton Trans. {\bf 50}, 13580 (2021). 
 
 \bibitem{Wang2017} H.-T. Wang, M. K. Srivastava, C.-C. Wu, S.-H. Hsieh, Y.-F. Wang, Y.-C. Shao, Y.-H. Liang, C.-H. Du, J.-W. Chiou, C.-M. Cheng, J.-L. Chen, C.-W. Pao, J.-F. Lee, C. N. Kuo, C. S. Lue, M.-K. Wu, W.-F. Pong 
Sci. Rep. {\bf 7},  40886 (201

 
\bibitem{Luo2016}C. W. Luo, P. C. Cheng, C. M. Tu, C. N. Kuo, C. M. Wang, and C. S. Lue, New J. Phys. \textbf{18}, 7 (2016).
 


\bibitem{Wang2015} L. M. Wang, Chih-Yi Wang, Guan-Min Chen,C. N. Kuo and C. S. Lue, 
New J. Phys. {\bf 17}, 033005 (2015).


\bibitem{mechanismCDW3413} A. F. Fang, X. B. Wang, P. Zheng, and
N. L. Wang, Phys. Rev. B \textbf{90}, 035115 (2014).




\bibitem{CNKuo2014}
C. N. Kuo, H. F. Liu, C. S. Lue, L. M. Wang, C. C. Chen, and Y. K. Kuo
Phys. Rev. B {\bf 89}, 094520 (2014).

\bibitem{ChemistryRareEarth}
Roman Gumeniuk, Chem. Rare Earths, {\bf 54}, Ch.304, 43 (2018).


\bibitem{CrystEngComm}
Iain W. H. Oswald, Binod K. Rai, Gregory T. McC
andless, Emilia Morosan and Julia Y. Chan,
Cryst. Eng. Comm. {\bf 19}, 3381 (2017).

\bibitem{JPhysCM}
J. P. A. Westerveld, D. M. R. Lo Cascioi, H. Bakkeri, B. 0. Loopstra, and K. Goubitz,
J. Phys.: Condens. Matter {\bf 1} 5689(1989).

\bibitem{Dalton2015}
R. Gumeniuk, M. Schöneich, K. O. Kvashnina, L. Akselrud, A. A. Tsirlin,
M. Nicklas, W. Schnelle, O. Janson, Q. Zheng, C. Curfs, U. Burkhardt, U. Schwarz,  and A. Leithe-Jasper, Dalton Trans. {\bf 44}, 5638 (2015).  

\bibitem{Suyama2018}K. Suyama, K. Iwasa, Y. Otomo, K. Tomiyasu, H. Sagayama, R. Sagayama, H. Nakao, R. Kumai, Y. Kitajima, F. Damay, J.-M. Mignot, A. Yamada, T. D. Matsuda, and Y. Aoki, Phys. Rev. B \textbf{97}, 23 (2018).

\bibitem{mechanisms} Xuetao Zhu, Jiandong Guo, Jiandi Zhang, and
E. W. Plummer, Adv. Phys. X \textbf{2}, 622 (2017).

\bibitem{Gabovich} A.~M.~Gabovich, A.~I.~Voitenko, and M.~Ausloos,Phys. Rep.- Rev. Sec. Phys. Lett. \textbf{367}, 583 (2002).

\bibitem{CDWSCcompetition} A. M. Gabovich, A. I. Voitenko, T. Ekino,
Mai Suan Li, H. Szymczak, and M. Pekala, Adv. Cond. Mat. Phys. \textbf{2010}, 681070 (2010).

\bibitem{YBCO} Maxime Leroux, Vivek Mishra, Jacob P. C. Ruff, Helmut
Claus, Matthew P. Smylie, Christine Opagiste, Pierre Rodi\'{E}re, Asghar
Kayani, G. D. Gu, John M. Tranquada, Wai-Kwong Kwok, Zahirul Islam,
and Ulrich Welp, Proc. Natl. Acad. Sci. USA \textbf{116}, 10691 (2019).


\bibitem{Cheung2018}Y. W. Cheung, Y. J. Hu, M. Imai, Y. Tanioku, H. Kanagawa, J. Murakawa, K. Moriyama, W. Zhang, K. T. Lai, K. Yoshimura, F. M. Grosche, K. Kaneko, S. Tsutsui, S. K. Goh, Phys. Rev. B. \textbf{98}, 161103 (2018).



\bibitem{Kiss2007} T. Kiss, T. Yokoya, A. Chainani, S. Shin, T. Hanaguri,
M. Nohara, H. Takagi, Nat. Phys. \textbf{3}, 720-725 (2007).


\bibitem{CDWphonon} Jianqiang Hou, Chi Ho Wong, Rolf Lortz, Romain
Sibille, and Michel Kenzelmann, Phys. Rev. B \textbf{93}, 134505 (2016).


\bibitem{Thermalconductivity} S. Y. Zhou, H. Zhang, X. C. Hong, B.
Y. Pan, X. Qiu, W. N. Dong, X. L. Li, and S. Y. Li, Phys. Rev. B \textbf{86}, 064504 (2012).



\bibitem{Kase2011} N. Kase, H. Hayamizu, and J. Akimitsu, Phys. Rev. B \textbf{83}, 184509 (2011).

\bibitem{Wang2012} Kefeng Wang and C. Petrovic, Phys. Rev B \textbf{86}, 024522 (2012).

\bibitem{Lue2016}C. S. Lue, C. N. Kuo, C. W. Tseng, K. K. Wu, Y.-H. Liang, C.-H. Du, and Y. K. Kuo, , Phys. Rev. B \textbf{93}, 245119 (2016).




\bibitem{Goh2} Wing Chi Yu, Yiu Wing Cheung, Paul J. Saines, Masaki
Imai, Takuya Matsumoto, Chishiro Michioka, Kazuyoshi Yoshimura, and
Swee K. Goh, Phys. Rev. Lett. \textbf{115}, 207003 (2015).

\bibitem{Luo2018} Jun Luo, Jie Yang, S Maeda, Zheng Li, and Guo-Qing
Zheng, Chin. Phys. B \textbf{27}, 077401 (2018).

\bibitem{swave} P. K. Biswas, Z. Guguchia, R. Khasanov, M. Chinotti, L. Li, Kefeng Wang, C. Petrovic, and E. Morenzon, Phys. Rev. B \textbf{92}, 195122 (2015).

\bibitem{Akimitsu}
Hiroki Hayamizu, Naoki Kase, and Jun Akimitsu, J. Phys. Soc. Jpn. {\bf 80},  SA114 (2011).






 
 
\bibitem{Biswas2014} P. K. Biswas, A. Amato, Kefeng Wang, C. Petrovic, R. Khasanov, H. Luetkens and E. Morenzoni, J. Phys.: Conf. Ser. \textbf{551}, 012029 (2014)
 

\bibitem{multibandmusR} P. K. Biswas, A. Amato, R. Khasanov, H. Luetkens,
Kefeng Wang, C. Petrovic, R. M. Cook, M. R. Lees, and E. Morenzoni, Phys. Rev. B \textbf{90}, 144505 (2014).

\bibitem{multibandNMR} R. Sarkar, F. Brückner, M. Ganther, Kefeng
Wang, C. Petrovic, P.K. Biswas, H. Luetkens, E. Morenzoni, A. Amato,
and H-H. Klauss, Physica B \textbf{479}, 51 (2015).
 

 
 \bibitem{Science} T.~Yokoya, T.~; Kiss, A.~Chainani, S.~Shin,
M.~Nohara, and H.~Takagi, Science \textbf{294}, 2518 (2001).

\bibitem{Boaknin} Etienne Boaknin, M. A. Tanatar, Johnpierre Paglione,
D. Hawthorn, F. Ronning, R. W. Hill, M. Sutherland, Louis Taillefer,
Jeff Sonier, S. M. Hayden, and J. W. Brill, Phys. Rev. Lett. \textbf{90}, 117003 (2003).


\bibitem{Prozorov} J. D. Fletcher, A. Carrington, P. Diener, P. Rodi\'{E}re,
J. P. Brison, R. Prozorov, T. Olheiser, and R. W. Giannetta, 
Phys. Rev. Lett. \textbf{98}, 057003 (2007).


 
 
 
 
 
 
 
\bibitem{FS} Xiaoye Chen, Swee K. Goh, David A. Tompsett, Wing Chi
Yu, Lina Klintberg, Sven Friedemann, Hong'En Tan, Jinhu Yang, Bin
Chen, M. Imai, Kazuyoshi Yoshimura, Monika B. Gamza, F. Malte Grosche,
and Michael L. Sutherland, Phys. Rev. B \textbf{93}, 235121 (2016).

\bibitem{BCS}J. Bardeen, L. N. Cooper, J. R. Schrieffer, Phys. Rev. \textbf{106}, 162–164 (1957).

\bibitem{Annett1996} J. F. Annett, N. Goldenfeld, A. J. Leggett, J. Low Temp. Phys. \textbf{105}, 473–482 (1996).

\bibitem{Sauls2022}M. Zarea, H. Ueki, J. A. Sauls, arXiv:2201.07403 (2022).

\bibitem{TimmonsPdTe2} E. I. Timmons, S. Teknowijoyo, M. Ko\'{n}czykowski,
O. Cavani, M. A. Tanatar, Sunil Ghimire, Kyuil Cho, Yongbin Lee, Liqin
Ke, Na Hyun Jo, S. L. Bud'ko, P. C. Canfield, Peter P. Orth, Mathias
S. Scheurer, and R. Prozorov, Phys. Rev. Res. \textbf{2}, 023140 (2020).

\bibitem{CaK1144} S. Teknowijoyo, K. Cho, M. Ko\'{n}czykowski, E.
I. Timmons, M. A. Tanatar, W. R. Meier, M. Xu, S. L. Bud'ko, P. C.
Canfield, and R. Prozorov, Phys. Rev. B \textbf{97}, 140508(R) (2018).





\bibitem{Anderson} P. Anderson, J. Phys. Chem. Solids \textbf{11}, 26 (1959).

\bibitem{AG} A. A. Abrikosov and L. P. Gor'kov, Zh. Eksp. Teor. Fiz. \textbf{35}, 1558 (1958) {[}Sov. Phys. JETP \textbf{8}, 1090 (1959){]}. A. A. Abrikosov and L. P. Gor'kov, Zh. Eksp. Teor. Fiz. \textbf{36}, 319 (1959) {[}Sov. Phys. JETP \textbf{9}, 220 (1959){]}.

\bibitem{Hohenberg1964} P. Hohenberg, Sov. Phys. JETP {\bf 18} 834 (1964).

\bibitem{GolubovMarzin} A. A. Golubov and I. I. Mazin
Phys. Rev. B \textbf{55}, 15146 (1997).

\bibitem{tccuprates} F. Rullier-Albenque, H. Alloul, and R. Tourbot,
Phys. Rev. Lett. \textbf{91}, 047001 (2003).

\bibitem{IBSTcsuppression} K. Cho, M. Ko\'{n}czykowski, S. Teknowijoyo,
M. A. Tanatar, and R. Prozorov, Supercond. Sci. Technol. \textbf{31},
064002 (2018).


\bibitem{Openov1997}L. A. Openov, J. Exp. Theor. Phys. Lett. \textbf{66},
10 (1997). 

\bibitem{Openov2004}L. A. Openov, Phys. Rev. B \textbf{69}, 22 (2004).

\bibitem{Cho2022}K. Cho, M. Ko\'{n}czykowski, S. Ghimire, M. A. Tanatar, L.-L. Wang, V. G. Kogan, R. Prozorov, Phys. Rev. B \textbf{105}, 24506 (2022).

\bibitem{DisorderSOCFu} K. Michaeli and L. Fu, Phys. Rev. Lett. \textbf{109},
187003 (2012).

\bibitem{OurDisorderSOC} M. S. Scheurer, M. Hoyer, and J. Schmalian,
Phys. Rev. B \textbf{92}, 014518 (2015).

\bibitem{BrydonScattering} D. C. Cavanagh and P. M. R. Brydon, Phys.
Rev. B \textbf{101}, 054509 (2020). 

\bibitem{Damask1963} A. C. Damask and G. J. Dienes, \textit{Point Defects
in Metals}, (Gordon \& Breach, London, U.K., 1963).

\bibitem{Thompson1969}M. W. Thompson, \textit{Defects and Radiation Damage
in Metals}, Cambridge Monographs on Physics (Cambridge University
Press, Cambridge, U.K., 1969).

\bibitem{Westerveld1989}J. P. A. Westerveld, D. M. R. L. Cascio, H. Bakker, B. O. Loopstra, K. Goubitz, J. Phys.: Conden. Matt. \textbf{1}, 5689–5702 (1989).

\bibitem{Westerveld1987} J. P. A. Westerveld, D. M. R. L. Cascio, H. Bakker, J. Phys. F: Met. Phys. \textbf{17} 1963 (1987). 



\bibitem{anisotropy} M. A. Tanatar, N. Ni, C. Martin, R. T. Gordon,
H. Kim, V. G. Kogan, G. D. Samolyuk, S. L. Bud'ko, P. C. Canfield,
and R. Prozorov, Phys. Rev. B \textbf{79}, 094507 (2009).

\bibitem{SUST} M. A. Tanatar, N. Ni, S. L. Bud'ko, P. C. Canfield,
and R. Prozorov, Supercond. Sci. Technol. \textbf{23}, 054002 (2010).

\bibitem{TimmonsRSI} E. I. Timmons, M. A. Tanatar, Yong Liu, Kyuil
Cho, T. A. Lograsso, M. Ko\'{n}czykowski, R. Prozorov, Rev. Sci. Instrum. \textbf{91}, 073904 (2020).

\bibitem{VanDegrift1975RSI}C. T. Van Degrift, Rev. Sci. Instrum. \textbf{46}, 599--607 (1975).
 
\bibitem{Prozorov2000PRB}R. Prozorov, R. W. Giannetta, A. Carrington,
F. M. Araujo-Moreira, Phys. Rev. B \textbf{62}, 115 (2000).

\bibitem{Prozorov2000APL}R. Prozorov, R. W. Giannetta, A. Carrington,
P. Fournier, R. L. Greene, P. Guptasarma, D. G. Hinks, A. R. Banks, Appl. Phys. Lett. \textbf{77}, 4202 (2000).

\bibitem{Prozorov2021}R. Prozorov, Phys. Rev. App. \textbf{16}, 24014 (2021).

\bibitem{Prozorov2006SST}R. Prozorov and R. W. Giannetta, Supercon. Sci. Techn. \textbf{19}, R41 (2006).
 
\bibitem{Prozorov2011RPP_review}R. Prozorov and V. G. Kogan, Rep. Prog. Phys. \textbf{74}, 124505 (2011). 
 

\bibitem{Giannetta2021}R. W. Giannetta, A. Carrington and R. Prozorov,
J. Low. Temp. Phys. in print (2021). arXiv:2109.07616 




\bibitem{BaKMathesson} R. Prozorov, M. Ko\'{n}czykowski,
M. A. Tanatar, H.-H. Wen, R. M. Fernandes and P. C. Canfield, npj Quantum Materials \textbf{4}, 34 (2019).


\bibitem{PRX} R. Prozorov, M. Ko\'{n}czykowski, M. A. Tanatar, A.
Thaler, S. L. Bud'ko, P. C. Canfield, V. Mishra, and P. J. Hirschfeld,
Phys. Rev. X \textbf{4}, 041032 (2014).


\bibitem{Slebarski2016}A. \'{S}lebarski, J. Goraus, M. M. Ma\'{s}ka, P. Witas, M. Fija\l{}kowski, C. T. Wolowiec, Y. Fang, and M. B. Maple, Phys. Rev. B \textbf{93}, 24 (2016).








\bibitem{Cedomirnpj} Lijun Li, Xiaoyu Deng, Zhen Wang, Yu Liu, Milinda Abeykoon, Eric Dooryhee, Aleksandra Tomic, Yanan Huang, John B. Warren, Emil S. Bozin, Simon J. L. Billinge, Yuping Sun, Yimei Zhu, Gabriel Kotliar \& Cedomir Petrovic, npj Quantum Materials, \textbf{2}, 11 (2017).

\bibitem{NMR2015} C. N. Kuo, C. W. Tseng, C. M. Wang, C. Y. Wang,
Y. R. Chen, L. M. Wang, C. F. Lin, K. K. Wu, Y. K. Kuo, and C. S.
Lue, Phys. Rev. B \textbf{91}, 165141 (2015).

\bibitem{Naito1982} M.~Naito and S.~Tanaka, J. Phys. Soc. Jpn. \textbf{51}, 219 (1982).


\bibitem{Naito2nd}M.~Naito and S.~Tanaka, 
J. Phys. Soc. Jpn. \textbf{51}, 228 (1982).


\bibitem{degiorgi}
M. Chinotti, J. Ethiraj, C. Mirri, Xiangde Zhu, Lijun Li, C. Petrovic, and L. Degiorgi
Phys. Rev. B {\bf 97}, 045117 (2018).



\bibitem{spindisorder} 
B.R. Coles, Adv. Phys. {\bf 7}, 40 (1958). 





\bibitem{Tinkham_book} M. Tinkham \emph{Introduction to Superconductivity}, 2nd ed., (Dover Publications, N.Y., 1996).





\bibitem{BaP} Y. Mizukami, Y. Ko\'{n}czykowski, M. Kawamoto, S. Kurata, S. Kasahara, K. Hashimoto, V. Mishra, Y. Kreisel, A. Wang, P. J. Hirschfeld, Y. Matsuda, and T. Shibauchi, Nat. Commun. \textbf{5}, 5657 (2014).

\bibitem{Prozorov2000}R. Prozorov, R. W. Giannetta, P. Fournier,
and R. L. Greene, Phys. Rev. Lett. \textbf{85}, 3700 (2000). 

\bibitem{Slebarski2018} A. \'{S}lebarski, P. Zajdel, M. Fija\l kowski, M. M. Ma\'{s}ka, P. Witas, J. Goraus, Y. Fang, D. C. Arnold, and M. B. Maple. New J. Phys. \textbf{20}, 103020 (2018).

\bibitem{SigristUeda} M. Sigrist and K. Ueda, Rev. Mod. Phys. \textbf{63}, 239 (1991).

\bibitem{Scheurer2016} M. S. Scheurer, \textit{Mechanism, symmetry
and topology of ordered phases in correlated systems}, Ph.D. thesis, Karlsruher Institut für Technologie (KIT) (2016).

\bibitem{Hoyer2015} M. Hoyer, M. S. Scheurer, S. V. Syzranov, and
J. Schmalian, Phys. Rev. B \textbf{91}, 054501 (2015).

\bibitem{ChubukovSPIS} S. Maiti and A. V. Chubukov, Phys.~Rev.~B
\textbf{87}, 144511 (2013).

\bibitem{LaNi} Arushi, D. Singh, A. D. Hillier, M. S. Scheurer, and
R. P. Singh, Phys.~Rev.~B \textbf{103}, 174502 (2021).

\bibitem{ScheurerPRB2016} M. S. Scheurer, Phys. Rev. B \textbf{93},
174509 (2016).

\bibitem{Brydon} P. M. R. Brydon, S. Das Sarma, H.-Y. Hui, and J.
D. Sau, Phys. Rev. B \textbf{90}, 184512 (2014).

\bibitem{TBGApplication} R. Samajdar and M. S. Scheurer, Phys. Rev.
B \textbf{102}, 064501 (2020).

\bibitem{Kogan2021}V. G. Kogan and R. Prozorov, Phys. Rev.
B 103, 054502 (2021). 

\bibitem{Kogan2009} V. G. Kogan, C. Martin, and R. Prozorov, Phys.
Rev. B 80, 14507 (2009). 

\bibitem{hodeau1982ssc} J.~L. Hodeau, M. Marezio, J.~P. Remeika,
and C.-H. Chen, 
Solid State Comm. \textbf{42} (2), 97 (1982).

\bibitem{WIEN2k} P.~Blaha, K.~Schwarz, G.~K.~H. Madsen, D.~Kvasnicka, J.~Luitz, R.~Laskowski, F.~Tran, and L.~D. Marks, \textit{WIEN2k, An Augmented Plane Wave + Local Orbitals Program for Calculating Crystal Properties} (Karlheinz Schwarz, Techn. Universität Wien, Austria, 2018).

\bibitem{perdew1996prl} J.~P. Perdew, K. Burke, and M. Ernzerhof,
Phys. Rev. Lett. \textbf{77}, 3865 (1996).


\bibitem{blochl1994prb} P.~E. Bl\"ochl, O. Jepsen, and O.~K. Andersen, Phys. Rev. B \textbf{49}, 16223 (1994).

\end{thebibliography}
\end{document}